\documentclass[aps, pre, reprint, superscriptaddress]{revtex4-2}

\usepackage{amssymb}
\usepackage{amsmath}
\usepackage{amsfonts}
\usepackage[colorlinks,linkcolor=red,urlcolor=blue,citecolor=blue]{hyperref}
\usepackage{graphics}
\usepackage{txfonts}
\usepackage{tikz}  
\usetikzlibrary{arrows,shapes,positioning,shadows,backgrounds,fit}
\usepackage{braket}

\newcommand{\Tr}{{\rm Tr}}
\newcommand{\dd}{{\rm d}}
\newcommand{\sg}{\hat{\sigma}}
\renewcommand{\a}{\hat{a}}
\newcommand{\adag}{\hat{a}^\dagger}
\newcommand{\A}{\hat{A}}
\renewcommand{\H}{\hat{H}}
\newcommand{\rrho}{\hat{\rho}}
\newcommand{\ttau}{\hat{\tau}}

\graphicspath{{figs/}}
\begin{document}

\title{Thermodynamics of precision in quantum nano-machines}
\author{Antoine Rignon-Bret}
\affiliation{School of Physics, Trinity College Dublin, College Green, Dublin 2, Ireland}
\affiliation{\'{E}cole Normale Sup\'{e}rieure, 45 rue d'Ulm, F-75230 Paris, France}
\author{Giacomo Guarnieri}
\affiliation{School of Physics, Trinity College Dublin, College Green, Dublin 2, Ireland}
\author{John Goold}
\affiliation{School of Physics, Trinity College Dublin, College Green, Dublin 2, Ireland}
\author{Mark T. Mitchison}
\email{mark.mitchison@tcd.ie}
\affiliation{School of Physics, Trinity College Dublin, College Green, Dublin 2, Ireland}
\begin{abstract}
Fluctuations strongly affect the dynamics and functionality of nanoscale thermal machines. Recent developments in stochastic thermodynamics have shown that fluctuations in many far-from-equilibrium systems are constrained by the rate of entropy production via so-called thermodynamic uncertainty relations. These relations imply that increasing the reliability or precision of an engine's power output comes at a greater thermodynamic cost. Here we study the thermodynamics of precision for small thermal machines in the quantum regime. In particular, we derive exact relations between the power, power fluctuations, and entropy production rate for several models of few-qubit engines (both autonomous and cyclic) that perform work on a quantised load. Depending on the context, we find that quantum coherence can either help or hinder where power fluctuations are concerned. We discuss design principles for reducing such fluctuations in quantum nano-machines, and propose an autonomous three-qubit engine whose power output for a given entropy production is more reliable than would be allowed by any classical Markovian model.
\end{abstract}

\maketitle

\section{Introduction}

Close examination of a small-scale system typically reveals significant fluctuations due to thermal noise. Not only do these fluctuations open a window on otherwise hidden phenomena~\cite{Einstein1905,Roldan2010,RibezziCrivellari2014}, they also exert a decisive influence on the functionality of nanoscale machines --- such as atomic~\cite{Rossnagel2016,Lindenfels2019,Horne2020} or molecular~\cite{Kolomeisky2007} motors. On a fundamental level, the occurrence of microscopic fluctuations is inextricably linked to dissipation and the emergence of macroscopic irreversibility, which originates from the same underlying randomness. Yet while the quantitative relation between fluctuations and dissipation has long been understood for equilibrium systems~\cite{Callen1951}, the principles that connect fluctuations to irreversible behaviour far from equilibrium are still in the process of being uncovered~\cite{Esposito2009,Campisi2011,Seifert2012,Haenggi2015,Landi2020}.

Recently, it was discovered~\cite{Barato2015,Gingrich2016} that non-equilibrium fluctuations are constrained by dissipation through a rather general class of inequalities known as thermodynamic uncertainty relations (TURs)~\cite{Horowitz2019}. Broadly speaking, TURs dictate that the currents which characterise any non-equilibrium scenario, e.g. the heat current powering an engine, must fluctuate by a certain minimum amount that is controlled by the rate of entropy production, such that reduced fluctuations necessitate increased entropy production. This principle has striking consequences for the performance of heat engines governed by classical physics: approaching the ultimate Carnot efficiency at finite average power output is possible only if the power fluctuations diverge~\cite{Campisi2016,Holubec2017,Pietzonka2018}. TURs also limit the accuracy of autonomous clocks~\cite{Barato2016,Erker2017,Mitchison2019,Pearson2020,Schwarzhans2020,Milburn2020,Woods2020} and biochemical sensors~\cite{Harvey2020}, and can be used to infer difficult-to-measure quantities such as the entropy production of molecular motors~\cite{Pietzonka2016,Seifert2019}.

Due to their fundamental and practical importance, a wealth of recent research has aimed at extending TURs beyond their original realms of validity, finding that they apply to classical stochastic systems under remarkably general conditions. Examples include finite observation times~\cite{Pietzonka2017,Horowitz2017}, discrete-time processes~\cite{Proesmans2017,Chiuchiu2018}, counting observables~\cite{Garrahan2017}, and feedback protocols~\cite{Potts2019}. Tighter bounds have been derived~\cite{Polettini2016,Falasco2020} and connections have been found with other important concepts of non-equilibrium thermodynamics, including fluctuation theorems~\cite{Timpanaro2019,Hasegawa2019a} and information theory~\cite{Dechant2018,Hasegawa2019,Dechant2020}. A substantial body of theoretical work has also been devoted to the thermodynamics of precision for quantum systems. In general, the classical TURs can be violated in the presence of quantum coherence, which can boost the reliability of nanoscale thermoelectric generators~\cite{Ptaszynski2018,Agarwalla2018,Brandner2018,Liu2019}. Quantum generalisations of TURs have been proved for steady-state~\cite{Guarnieri2019} and cyclic~\cite{Miller2020} quantum heat engines, for quantum systems under linear-response conditions~\cite{Macieszczak2018}, and for Markovian open quantum systems subjected to continuous weak measurements~\cite{Federico2019,Hasegawa2020prl,Hasegawa2020}. While these general bounds are useful for understanding fundamental limits, the precise relation between entropy production and non-equilibrium fluctuations must be assessed for each specific system on a case-by-case basis~\cite{Segal2018,Saryal2019,Buffoni2020,Benenti2020}.

Here, we analyse the thermodynamics of precision for small quantum heat engines that perform work on a load with an infinite-dimensional Hilbert space. Our study is motivated by recent experimental implementations of nanoscale devices whose work output is stored in the vibrations of a mechanical oscillator~\cite{Lindenfels2019,Wen2019,Horne2020}. At such small scales, energetic fluctuations are unavoidable and may significantly affect performance~\cite{Tonner2005,Boukobza2006, Youssef2009, Brunner2012,GelbwaserKlimovsky2013,Mari2015,Levy2016, Niedenzu2019}. We show this explicitly by deriving exact equalities connecting the power, its fluctuations, and the rate of entropy production for some basic models of quantum heat engines that have been considered in the literature. We will henceforth refer to these equalities as \textit{TUR ratios}, since they share the structure and spirit of the TURs.

We begin in Sec.~\ref{sec:2QE} with the two-qubit autonomous engine introduced by Brunner et al.~\cite{Brunner2012}, which forms a minimal template for all quantum absorption machines~\cite{Linden2010,Mitchison2019}. We derive TUR ratios describing the power fluctuations of this two-qubit engine and discuss the underlying physical mechanisms that give rise to them. In Sec.~\ref{sec:3QE}, we use these insights to design a three-qubit autonomous engine that exhibits substantially reduced fluctuations due to quantum-coherent energy transport. However, coherence is not always helpful, as we show in Sec.~\ref{sec:flywheel} by considering the work output of a cyclic Otto engine with a qubit working medium and a harmonic-oscillator load~\cite{Lindenfels2019}. In that case, local coherence in the load's energy eigenbasis lead to increased energetic fluctuations compared to the analogous classical process. Our results reveal general design principles for reducing fluctuations in quantum nano-machines, and contribute towards a deeper understanding of the differences between quantum and classical heat engines. Units where $\hbar=k_B=1$ are used throughout. 

\section{Minimal autonomous two-qubit engine}
\label{sec:2QE}

\subsection{Two-qubit engine model}
\label{sec:2QE_model}

\begin{figure}
    \centering
    \includegraphics[width=0.7\linewidth]{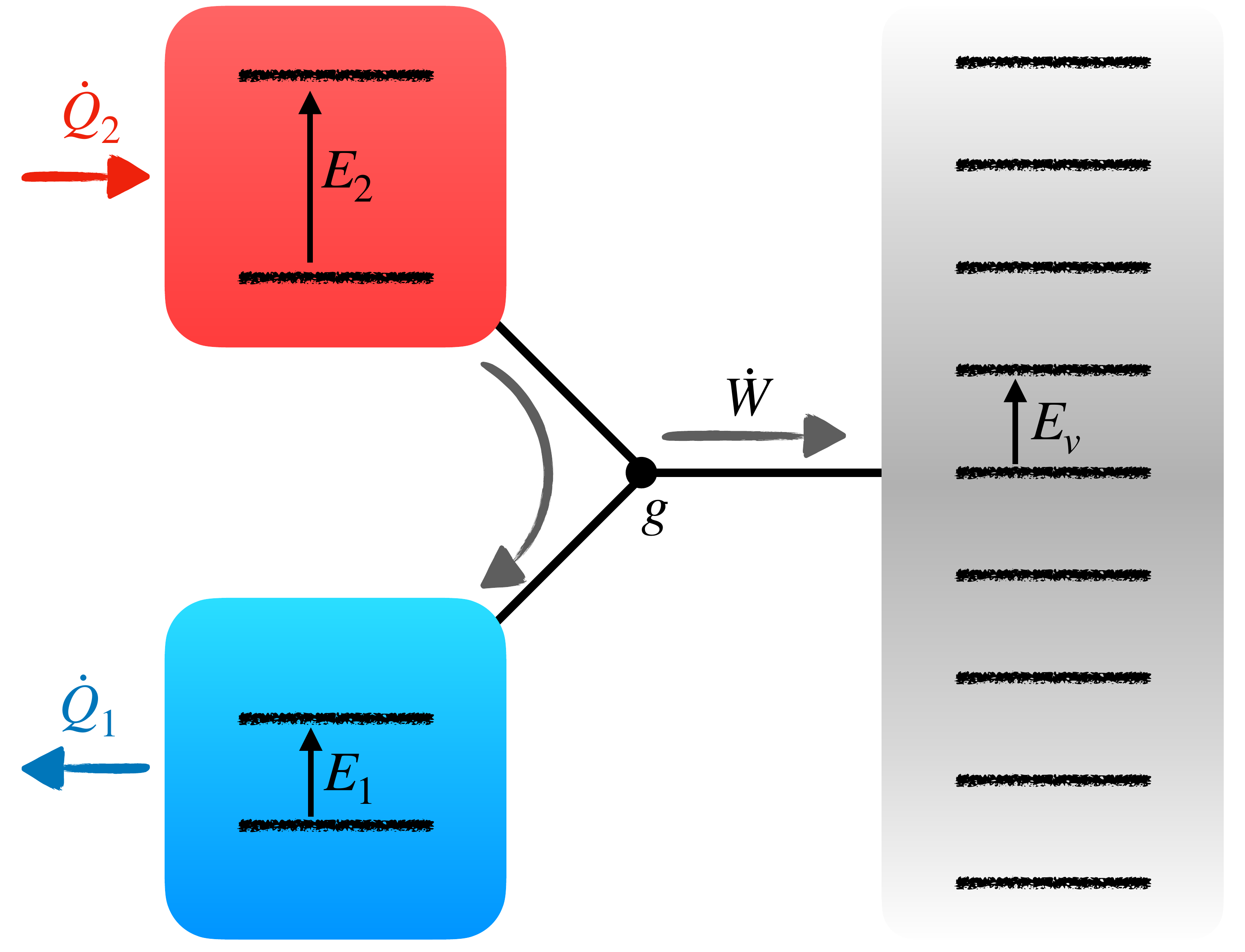}
    \caption{Illustration of an autonomous two-qubit engine coupled to an infinite-dimensional load via a tripartite interaction. The system is maintained out of equilibrium by heat currents $\dot{Q}_1$ and $\dot{Q}_2$ exchanged with thermal reservoirs at temperatures $T_1$ and $T_2>T_1$. A portion of the energy flowing from the hot qubit (red) to the cold qubit (blue) is diverted to peform work on the load (grey).}
    \label{fig:2QE}
\end{figure}

Our first example is the minimal model of an autonomous quantum heat engine introduced in Ref.~\cite{Brunner2012}, which is depicted schematically in Fig.~\ref{fig:2QE}. The machine consists of two qubits with energy splitting $E_1$ and $E_2$, coupled to independent heat reservoirs at the respective temperatures $T_1$ and $T_2$, where $T_2 > T_1$. This two-qubit engine performs work on a load system which is described by an infinite-dimensional ladder of equidistant energy eigenstates separated by an energy $E_v = E_2 - E_1$. The Hamiltonian of the system is given by $\H = \H_0 + \H_{\rm int}$, with
\begin{align}
    \label{2QE_H0}
    \H_0 &= \frac{1}{2}\sum_{j=1,2} E_j \sg^z_j + \hat{W}, \\
    \label{2QE_Hint}
    \H_{\rm int} & = g \left( \sg^+_1\sg^-_2\A^\dagger + \sg^-_1\sg^+_2\A  \right),
\end{align}
where $\sg_j^{x,y,z}$ denotes the standard Pauli operators describing qubit $j=1,2$, $\sg^\pm_j = (\sg^x_j \pm i \sg_j^y)/2$ are the spin lowering and raising operators, and we defined the load's energy operator $\hat{W} = \sum_{n=-\infty}^\infty n E_v \ket{n}_w\bra{n}$ and lowering operator ${\A =  \sum_{n=-\infty}^\infty \ket{n-1}_w\bra{n}}$. Similar tripartite engine models have been proposed in the context of quantum optics~\cite{Youssef2009} and optomechanics~\cite{Mari2015}.

The form of the interaction in Eq.~\eqref{2QE_Hint} allows energy quanta to flow from the hot qubit to the cold one, but only by simultaneously transferring quanta of energy $E_v = E_2-E_1$ to the load. This process is more likely than its time-reverse (whereby the load loses energy) so long as entropy is produced in accordance with the second law of thermodynamics. Each quantum of energy transferred to the load leads to entropy changes $\Delta S_1 = E_1/T_1$ and $\Delta S_2 =  -E_2/T_2$ associated with the heat exchanged with the cold and hot reservoirs, respectively. Denoting the total entropy production of this process by $\chi$, the engine's operating condition is then given by
\begin{equation}
    \label{positive_entropy_production}
    \chi \equiv \frac{E_1}{T_1} - \frac{E_2}{T_2} \geq 0.
\end{equation}

\begin{figure}[b]
    \centering
    \includegraphics[width=0.65\linewidth]{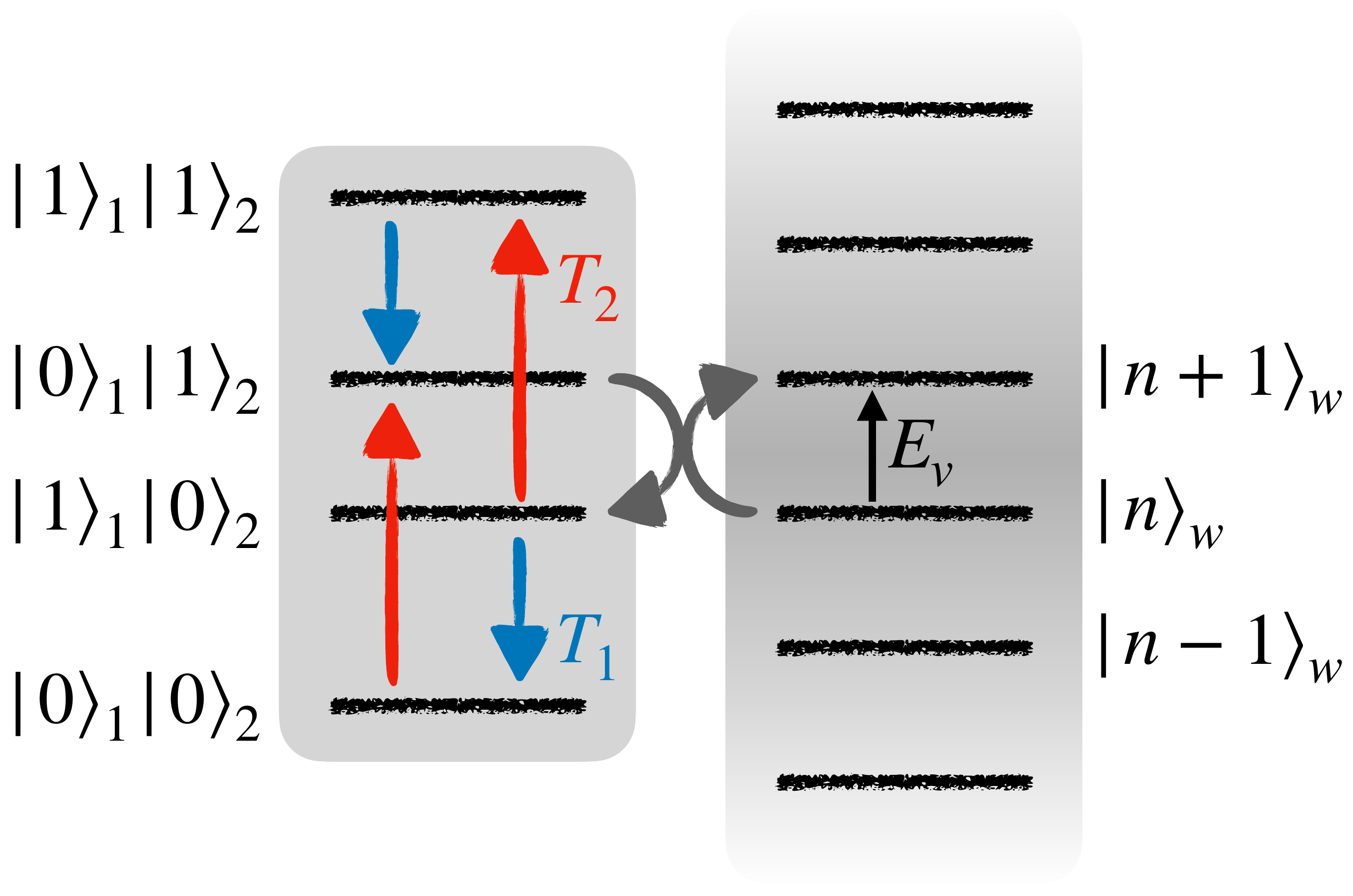}
    \caption{The virtual qubit is the pair of states $\{\ket{0}_1\ket{1}_2,\ket{1}_1\ket{0}_2\}$ in the composite Hilbert space of the two engine qubits. Thermal baths drive transitions between the four engine eigenstates, generating population inversion in the virtual qubit, i.e.~a negative temperature. Resonant coupling to the virtual qubit causes the load to thermalise with the negative virtual temperature, thus performing work.}
    \label{fig:virtual_qubit}
\end{figure}

To understand this condition microscopically, we identify a \textit{virtual qubit} in the composite Hilbert space of the hot and cold qubits~\cite{Brunner2012}. The virtual qubit is defined by the pair of states $\ket{0}_v = \ket{1}_1\ket{0}_2$ and $\ket{1}_v = \ket{0}_1\ket{1}_2$, where $\{\ket{0}_j,\ket{1}_j\}$ are the eigenstates of $\sg^z_j$. These virtual qubit states are eigenstates of $\H_0$ with energy splitting $E_v$, which exchange energy with the load via $\H_{\rm int}$ (see Fig.~\ref{fig:virtual_qubit}). In the absence of coupling to the load ($g=0$), the machine qubits are in equilibrium with their corresponding baths, so that the population of each energy eigenstate obeys the Boltzmann distribution, i.e.\ $P(\ket{1}_j)/P(\ket{0}_j) = e^{-E_j/T_j}$. It follows that the virtual qubit states are populated in the ratio $P(\ket{1}_v)/P(\ket{0}_v) = e^\chi = e^{- E_v/T_v} $, where $T_v = -E_v/\chi$ is the virtual temperature~\cite{Skrzypczyk2015}. When $\chi$ is large and positive, the virtual temperature is small and negative and the virtual qubit populations are inverted, i.e.\ $P(\ket{1}_v) \gg P(\ket{0}_v)$. Turning on a weak interaction $g\neq 0$ couples the virtual qubit to the load by driving transitions of the form $\ket{1}_v \ket{n}_w\leftrightarrow \ket{0}_v\ket{n+1}_w$. The population inversion of the virtual qubit biases the forward transition in favour of the reverse one, causing the load's energy to increase over time. In this way, the engine's operation can be understood as ``thermalisation'' of the load but with a negative virtual temperature. 

\subsection{Thermodynamics of precision for the two-qubit engine}
\label{sec:TUR_2QE_reset}

We now turn to the explicit description of the engine's dynamics. Following Refs.~\cite{Linden2010,Brunner2012}, we model the thermalising effect of the reservoirs by assuming that in each small time interval $\dd t$, the qubits either evolve coherently under the Hamiltonian $\H$ or one of them is randomly reset to a local thermal state $\ttau_j = e^{-\beta_j E_j\sg^z_j/2}/\mathcal{Z}_j$, with $\beta_j = 1/T_j$ and $\mathcal{Z}_j = \Tr\left( e^{-\beta_j E_j\sg^z_j/2}\right)$. For each qubit, this resetting is assumed to occur with probability $p$ per unit time and is described by the map $\rrho \to \ttau_j\otimes \Tr_j(\rrho)$. Taking the limit $\dd t\to 0$, the corresponding ensemble dynamics is given by the master equation
\begin{align}
    \label{reset_master_equation_2QE}
    \frac{\dd\rrho}{\dd t} & = -i [\H, \rrho] + \sum_{j=1,2} p \left( \ttau_j \otimes \Tr_j(\rrho)  - \rrho \right) \\
    \label{reset_ME_Lindblad}
    & = -i [\H, \rrho] + \sum_{j=1,2}\left(\gamma^+_j\mathcal{D}[\sg^+_j]  + \gamma^-_j\mathcal{D}[\sg_j^-] +  \gamma^z \mathcal{D}[\sg^z_j] \right)\rrho.
\end{align}
On the second line, the reset master equation is expressed explicitly in Lindblad form, where the dissipation super-operator is defined by $\mathcal{D}[\hat{L}]\bullet  = \hat{L}\bullet \hat{L}^\dagger - \tfrac{1}{2}\{\hat{L}^\dagger\hat{L},\bullet\}$, the gain and decay rates are $\gamma_j^\pm = p e^{\mp\beta_j E_j/2}/\mathcal{Z}_j$, and $\gamma^z = p/4$ is an effective local dephasing rate. The description of thermalisation in terms of local processes implicitly assumes that the coupling strength $g$ is not much larger than other energy or frequency scales, in particular $E_j$, $T_j$ and $p$. Note that, since $[\hat{H}_0,\hat{H}_{\rm int}] =0$, this local description of dissipation is thermodynamically consistent~\cite{Barra2015,Barra2018,DeChiara2018} and the spurious violations of thermodynamic laws predicted for systems with non-resonant interactions~\cite{Levy2014,Stockburger2016} do not arise.

The useful output of the two-qubit heat engine is quantified by the energy transferred to the load. This energy transfer is stochastic due to the inevitable fluctuations induced by the coupling to thermal reservoirs. We denote the mean energy of the load by $W = \langle \hat{W}\rangle$ and its variance by $\Delta_W = \langle \hat{W}^2\rangle - W^2$. In the long-time limit, the machine reaches a non-equilibrium steady state (NESS) characterised by stationary power $\dot{W} = \dd W/\dd t$ and power fluctuations $\dot{\Delta}_W=\dd \Delta_W/\dd t$. Our aim is to relate this energetic output with the rate of entropy production,
\begin{equation}
    \label{entropy_production_rate}
    \dot{\Sigma} = -\beta_1 \dot{Q}_1 - \beta_2 \dot{Q}_2,
\end{equation}
where $\dot{Q}_j = \Tr[\H\mathcal{D}_j \rrho]$ is the heat current entering the system from bath $j$ and $\mathcal{D}_j\rrho = p \left( \ttau_j \otimes \Tr_j(\rrho)  - \rrho \right)$ is the corresponding dissipator. Note that, since the fluctuating energy transfer to the load is identified with work output, we quantify entropy production in terms of the heat flux only and ignore the additional contribution associated with the load's growing von Neumann entropy~\cite{Spohn1978,Landi2020}.

To relate the machine's fluctuating power output to the entropy production, we exploit the exact solution of the master equation derived in Ref.~\cite{Brunner2012}, which is briefly described here and detailed fully in Appendix~\ref{app:2QE}. The state of the virtual qubit is characterised by the following two observables:
\begin{align}
    \label{bias_virtual}
    \hat{Z} &  =
    \ket{1}_v\bra{1} - \ket{0}_v \bra{0}= \tfrac{1}{2}\left(\sg_2^z - \sg_1^z\right), \\
    \label{occupation_virtual}
    \hat{N} & = \ket{1}_v\bra{1} + \ket{0}_v \bra{0} = \tfrac{1}{2}\left(1 - \sg_1^z\sg_2^z\right).
\end{align}
Respectively, these yield the mean bias (population inversion) $\langle \hat{Z}\rangle$ and the mean occupation $\langle\hat{N}\rangle$ of the virtual qubit states. In the absence of coupling to the load, these quantities take the values $\langle \hat{Z}\rangle_{\rm eq} \equiv \Tr(\hat{Z}\ttau_1\otimes\ttau_2) $ and $\langle \hat{N}\rangle_{\rm eq} \equiv \Tr(\hat{N}\ttau_1\otimes\ttau_2)$, where
\begin{align}
    \label{bias_mean}
    \langle \hat{Z}\rangle_{\rm eq} & = -\tanh(\beta_v E_v/2) \langle \hat{N}\rangle_{\rm eq}, \\
    \label{occupation_mean}
    \langle \hat{N}\rangle_{\rm eq}  & = \frac{1}{2}\left [1-\tanh(\beta_1E_1/2) \tanh(\beta_2E_2/2)\right],
\end{align}
which describe a qubit in equilibrium at inverse temperature $\beta_v = 1/T_v$ and with a total normalisation $\langle \hat{N}\rangle_{\rm eq}$ less than unity (because the qubit is virtual). For finite $g$, the virtual qubit drives a coherent current to the load described by the dimensionless operator  
\begin{equation}
    \label{current_operator}
    \hat{C} = i\left( \sg^-_1\sg^+_2\A - \sg^+_1\sg^-_2\A^\dagger \right).
\end{equation}
The average power delivered to the load is then $\dot{W} = g E_v\langle \hat{C}\rangle$. In the NESS, all currents are proportional to the bias, $\dot{Q}_j \propto \dot{W}\propto \langle \hat{C}\rangle \propto \langle \hat{Z}\rangle$, while the load's energetic fluctuations are determined by both $\langle \hat{Z}\rangle$ and $\langle \hat{N}\rangle$. Explicitly, we obtain the asymptotic solutions
\begin{align}
\label{work_2QE}
    \dot{W} & = \Gamma_2 E_v \langle \hat{Z}\rangle_{\rm eq}, \\
    \label{ent_2QE}
    \dot{\Sigma} & = \Gamma_2 \chi \langle \hat{Z}\rangle_{\rm eq}, \\
    \label{fluct_2QE}
    \dot{\Delta}_W & = \Gamma_2 E_v^2 \left[ \langle \hat{N}\rangle_{\rm eq} - \frac{2\Gamma_2(2p^2+g^2)}{p(p^2+2g^2)} \langle \hat{Z}\rangle_{\rm eq}^2 \right],
\end{align}
where $\Gamma_2 = g^2p/(p^2+2g^2)$ is the characteristic rate of energy transfer for the two-qubit engine. Due to the proportionality of the power and heat currents, the engine's efficiency is given by 
\begin{equation}
    \label{Otto_efficiency}
    \eta = \frac{\dot{W}}{\dot{Q}_2} = 1 - \frac{E_1}{E_2}.
\end{equation}
The Carnot bound, $\eta \leq \eta_C = 1 - T_1/T_2$, follows directly from condition~\eqref{positive_entropy_production}.

It is now straightforward to derive the TUR ratio
\begin{equation}
\label{TUR_ratio_2QE}
    \frac{\dot{\Delta}_W}{\dot{W}^2}\dot{\Sigma} = \chi \left[ \coth(\chi/2) - \frac{2\Gamma_2(2p^2+g^2)}{p(p^2+2g^2)} \langle\hat{Z}\rangle_{\rm eq}\right] \geq 2.
    \end{equation}
This relation encapsulates the trade-off between the precision of work deposition and its associated thermodynamic cost. For any given set of finite temperatures and qubit energies, Eq.~\eqref{TUR_ratio_2QE} gives the necessary conditions to minimise fluctuations in the engine's power output at fixed entropy production. The optimal operating point in this respect is $g=p$ (see Appendix~\ref{app:bounds}), which is effectively an impedance-matching condition between the bath-engine and engine-load couplings.

The inequality~\eqref{TUR_ratio_2QE}, which is proved in Appendix~\ref{app:bounds}, is of the same form as the classical TUR for steady-state currents~\cite{Barato2015,Gingrich2016}. We emphasise, however, that the conditions for the validity of the classical TUR do not hold here due to the presence of quantum coherences. To appreciate the importance of the bound~\eqref{TUR_ratio_2QE} for engine performance, we follow Ref.~\cite{Pietzonka2017} and rewrite it as
\begin{equation}
    \label{efficiency_power_tradeoff}
 \dot{\Delta}_W \geq  \frac{2T_1 \eta}{\eta_C - \eta}\dot{W},
\end{equation}
where we have used the steady-state energy balance equation $\dot{W} = \dot{Q}_1 + \dot{Q}_2$ to express the entropy production rate~\eqref{entropy_production_rate} in terms of the efficiency~\eqref{Otto_efficiency}. The above inequality implies that approaching the Carnot efficiency at finite power is possible only by allowing the fluctuations to diverge. For the two-qubit engine, the Carnot point corresponds to $\chi \to 0$, which is the limit where the bound in Eq.~\eqref{TUR_ratio_2QE} can be saturated. The virtual qubit's bias $\langle \hat{Z}\rangle_{\rm eq} \propto \tanh(\chi/2)$ tends to zero in this limit but its normalisation $\langle\hat{N}\rangle_{\rm eq}$ does not. Therefore, when the two-qubit engine operates at Carnot efficiency, its power output is zero on average but has non-zero fluctuations. We note that this behaviour was already discussed by Brunner et al.~\cite{Brunner2012} in terms of the qualitative notion of the ``strength of work''. Our results~\eqref{TUR_ratio_2QE} and~\eqref{efficiency_power_tradeoff} give this notion a clear quantitative meaning. 

The emergence of a TUR-like bound for this small quantum engine is not surprising, since the load's behaviour has some similarities with a classical stochastic process. In particular, the variance of the load's energy grows linearly in time, as expected for diffusive dynamics. Indeed, for $g\ll p$, the load's evolution can be accurately approximated by a classical random walk~\cite{Erker2017,Mitchison2019}, with probabilities per unit time of an upward or downward step related by $p_{\uparrow} = e^\chi p_\downarrow$. In this limit, the second term in Eq.~\eqref{TUR_ratio_2QE} can be neglected and one recovers the TUR ratio $\dot{\Delta}_W\dot{\Sigma}/\dot{W}^2 = \chi \coth(\chi/2)$ for the biased random walk, which is the prototypical model first used by Barato and Seifert~\cite{Barato2015} to illustrate the thermodynamics of precision. 

Conversely, at finite coupling $g$, the fluctuations are reduced relative to the classical random-walk case. To clarify this, it is instructive to rewrite Eq.~\eqref{TUR_ratio_2QE} as
\begin{equation}
    \label{TUR_ratio_coherence}
 \frac{\dot{\Delta}_W}{\dot{W}^2}\dot{\Sigma} = \chi\coth(\chi/2) \frac{\langle \hat{N}\rangle - 3\langle \hat{C}\rangle^2}{\langle \hat{N}\rangle_{\rm eq}}.
\end{equation}
 This form highlights two different ways in which a coherent coupling between the virtual qubit and the load reduces the TUR ratio. First, the coupling depletes the occupation of the virtual qubit below its equilibrium value, $\langle \hat{N}\rangle < \langle \hat{N}\rangle_{\rm eq}$. Considering the virtual qubit as an effective heat reservoir at temperature $T_v$, this depletion occurs because the reservoir is not macroscopic and thus experiences strong back-action from its coupling to the system. Second, the coupling reduces the TUR ratio by establishing a finite current $\langle \hat{C}\rangle$, which is associated with quantum coherences in the energy eigenbasis of the coupled qubit-load system~\cite{Mitchison2018}. Eq.~\eqref{TUR_ratio_coherence} thus suggests two ways in which small thermal machines might overcome the constraints of the classical TUR: either by harnessing small or non-Markovian heat sources, or by exploiting quantum coherences.

\subsection{Effect of local dephasing}
\label{sec:2QE_local_lindblad}

Having established the connection between power, fluctuations and dissipation, it is natural to ask whether a more propitious relationship between these quantities can be arranged. As shown by Eq.~\eqref{TUR_ratio_coherence}, the presence of quantum coherence can reduce the power fluctuations for a given entropy production rate. We now show that this effect is enhanced by considering a different dissipation model, in which the local dephasing terms proportional to $\gamma^z$ in Eq.~\eqref{reset_master_equation_2QE} are neglected. We thus consider the master equation
\begin{equation}
        \label{local_ME_Lindblad_2QE}
    \frac{\dd \hat{\rho}}{\dd t} = -i [\H, \rrho] + \sum_{j=1,2}\left(\gamma^+_j\mathcal{D}[\sg^+_j]  + \gamma^-_j\mathcal{D}[\sg_j^-] \right)\rrho.
\end{equation}
A local Lindblad equation of this form can be derived from a time-independent system-reservoir interaction under the assumption of weak coupling $g$ relative to the temperatures and the local qubit and load energies~\cite{Hofer2017}. In order to facilitate comparison with the reset model of Sec.~\ref{sec:TUR_2QE_reset}, we continue to parametrise the gain and loss rates as $\gamma_j^\pm = p e^{\mp\beta_j E_j/2}/\mathcal{Z}_j$.

\begin{figure}
    \centering
    \includegraphics[width=\linewidth]{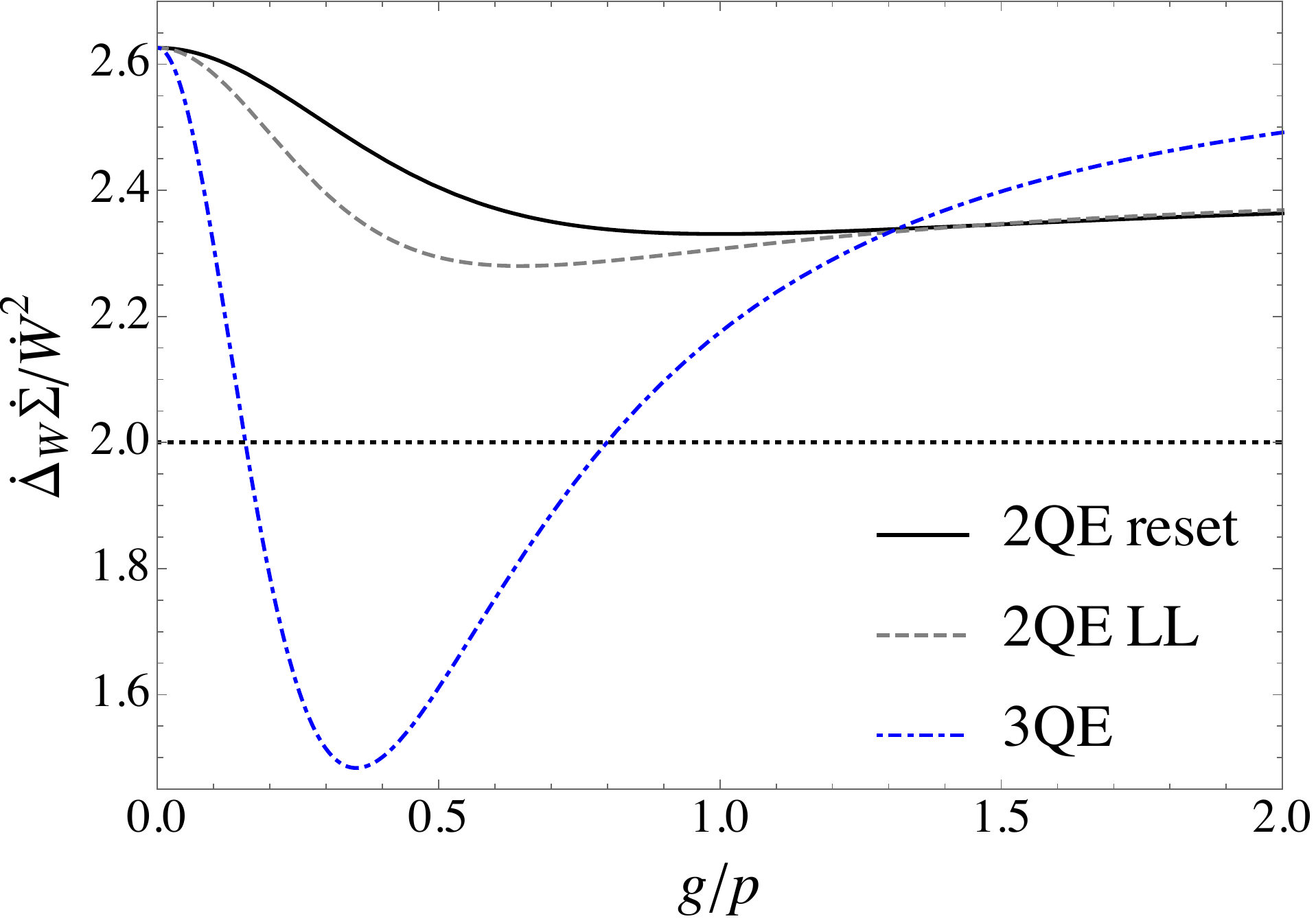}
    \caption{Thermodynamics of precision for autonomous few-qubit engines. The product of relative power fluctuations and entropy production rate is plotted as a function of the coherent coupling $g$ relative to the average dissipation rate $p$. Results are shown for the two-qubit engine (2QE) with a reset (solid black line) or local Lindblad (dashed grey line) thermalisation model, and for the three-qubit engine (3QE) model (dot-dashed blue line). The classical Markovian TUR bound is shown by the dotted black line. The parameters are $\beta_1E_1 = 3$ and $\beta_2 E_2 = 1$.
    \label{fig:TUR_ratio}}
\end{figure}

The dynamics is solved to find the NESS corresponding to Eq.~\eqref{local_ME_Lindblad_2QE} in Appendix~\ref{app:2QE}. The diagonal elements of the density matrix in the computational basis are found to obey identical equations of motion to the reset model of Eq.~\eqref{reset_master_equation_2QE}, while the off-diagonal elements experience reduced decoherence rates due to the absence of a $\gamma_z$ term. Explicitly, we find the solutions
\begin{align}
\label{work_LL}
    \dot{W} & = \Gamma_2' E_v \langle \hat{Z}\rangle_{\rm eq}, \\
    \label{ent_LL}
    \dot{\Sigma} & = \Gamma_2' \chi \langle \hat{Z}\rangle_{\rm eq}, \\
    \label{fluct_LL}
    \dot{\Delta}_W & = \Gamma_2' E_v^2 \left[ \langle \hat{N}\rangle_{\rm eq} - \frac{\Gamma_2'(5p^2+4g^2)}{p(p^2+4g^2)} \langle \hat{Z}\rangle_{\rm eq}^2 \right].
\end{align}
The results are very similar to the reset model, Eqs.~\eqref{work_2QE}--\eqref{fluct_2QE}, but with a modified characteristic rate of energy flux, $\Gamma_2' = 2g^2p/(p^2+4g^2)$, and somewhat reduced fluctuations. The TUR ratio then follows as
\begin{equation}
    \label{TUR_ratio_LL}
    \frac{\dot{\Delta}_W}{\dot{W}^2}\dot{\Sigma} = \chi \left[\coth(\chi/2) - \frac{\Gamma_2'(5p^2+4g^2)}{p(p^2+4g^2)}\langle\hat{Z}\rangle_{\rm eq}\right].
\end{equation}
Due to its different dependence on the coefficients $p$ and $g$, this quantity can take smaller values than the TUR ratio~\eqref{TUR_ratio_2QE} for the reset model, given the same temperatures and qubit energies. We show in Appendix~\ref{app:3QE_ME} that the bound $\dot{\Delta}_W\dot{\Sigma}/\dot{W}^2 \geq 1.982\ldots$ holds, which allows for fluctuations below the classical TUR. However, we have found that such violations are typically extremely small and occur only in a very limited region of the parameter space where $\beta_j E_j \ll 1$. The TUR ratios for the different 2QE models are compared in Fig.~\ref{fig:TUR_ratio}. 

\section{Autonomous Three-Qubit Engine}
\label{sec:3QE}

\subsection{Three-qubit engine model}
\label{sec:3QE_model}

\begin{figure}
    \centering
    \includegraphics[width=\linewidth]{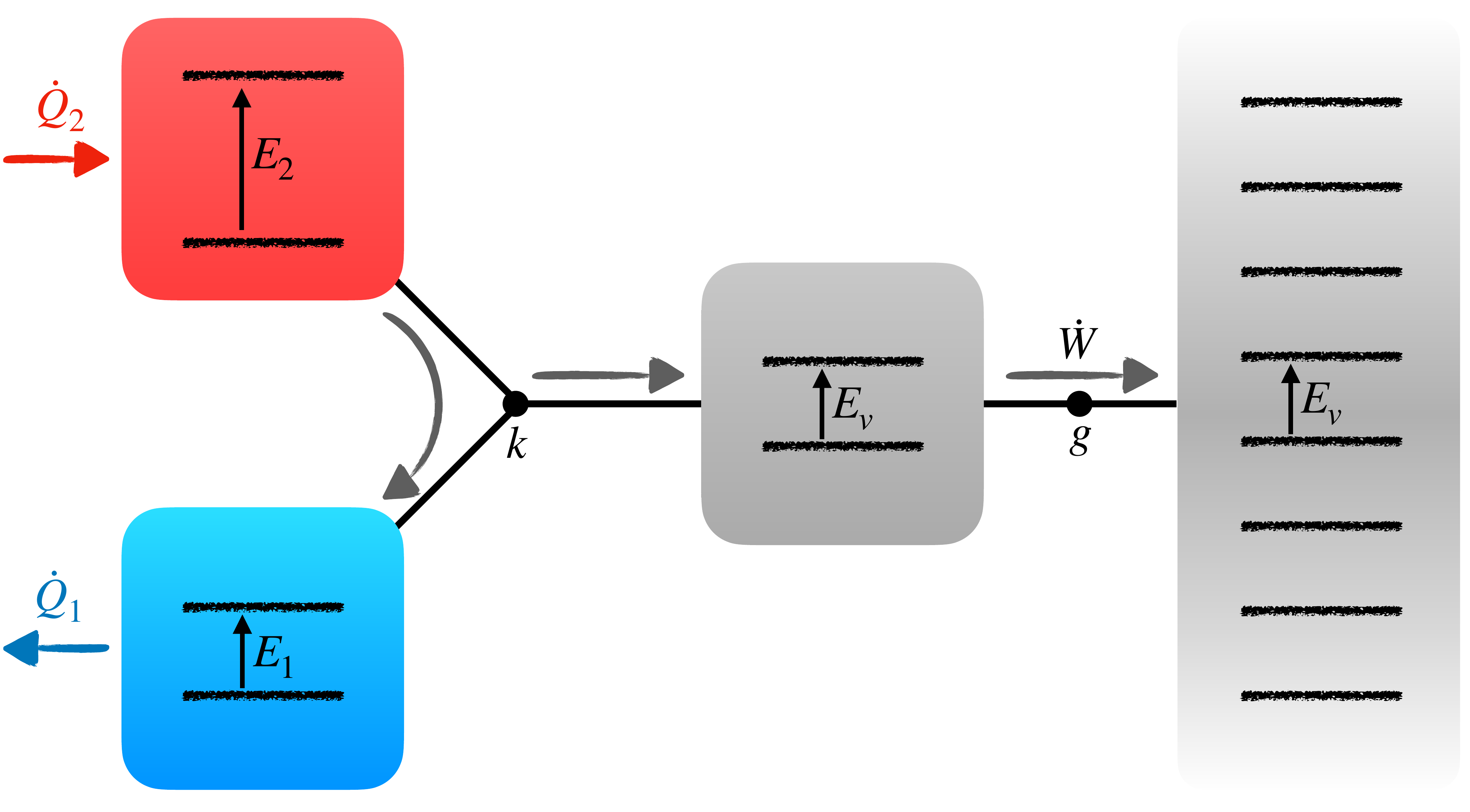}
    \caption{Schematic of the autonomous three-qubit engine. The energy flow from the baths is intermediated by an additional qubit resonant with the load. This boosts quantum coherence associated with energy transport, allowing for reduced relative power fluctuations.}
    \label{fig:3QE}
\end{figure}

As we have seen, fluctuations in the power output may be reduced in the presence of coherent energy transport, in accordance with previous studies~\cite{Agarwalla2018,Ptaszynski2018,Brandner2018}. We now use this insight to design an autonomous quantum nano-machine with a more reliable power output than would be classically allowed by the TUR. In particular, we modify the two-qubit engine by adding a third qubit that intermediates the flow of energy from the baths to the load, as depicted in Fig.~\ref{fig:3QE}. This shifts the coherent coupling between engine and load further away from the decohering effect of the thermal baths, thereby reducing power fluctuations. 

The Hamiltonian of the model is $\H = \H'_0 + \hat{V} + \hat{H}_{\rm int}$, where
\begin{align}
    \label{3QE_h0}
    \H'_0 & = \frac{1}{2}\sum_{j=1}^3 E_j \sg_j^z + \hat{W}, \\
    \label{3QE_V}
    \hat{V} & = k \left( \sg_1^+ \sg_2^- \sg_3^+ + \sg_1^- \sg_2^+ \sg_3^- \right), \\
    \label{3QE_Hint}
    \H_{\rm int} & = g \left( \sg_3^- \hat{A}^\dagger + \sg_3^+ \hat{A}\right),
\end{align}
with $\H'_0$ the free Hamiltonian, $\hat{V}$ the interaction between the three qubits, and $\H_{\rm int}$ the engine's coupling to the load. We take $E_3 = E_2 - E_1 = E_v$ to ensure that all interactions are resonant. As before, we assume that qubits 1 and 2 are locally coupled to thermal baths at temperatures $T_1$ and $T_2$, respectively. For $g=0$, the additional qubit thus thermalises to the virtual temperature, with bias given by
\begin{equation}
\label{bias_qubit3}
    \langle \sg_3^z\rangle_{\rm eq} = -\tanh(\beta_v E_v/2).
\end{equation}
This is analogous to Eq.~\eqref{bias_mean} but with unit normalisation (i.e.~$\langle \hat{N}\rangle = 1$) because the qubit is physical, not virtual.

To model the dynamics for finite $g$, we simplify the problem by assuming that the local thermalisation rate is much larger than both the coherent couplings $k$ and $g$. As a result, the rapidly damped qubits behave approximately like memoryless thermal reservoirs on the slow timescale over which energy is transported between the engine and the load. As shown in Appendix~\ref{app:3QE}, these qubits can be perturbatively eliminated under the Born-Markov approximation. The result is an effective master equation describing the joint state of qubit 3 and the load, which reads as
\begin{equation}
    \label{3QE_master_equation}
    \frac{\dd \hat{\rho}}{\dd t} = -i [\H_0 + \H_{\rm int}, \rrho] + \gamma^+\mathcal{D}[\sg^+_3]\rrho  + \gamma^-\mathcal{D}[\sg_3^-] \rrho,
\end{equation}
where $\H_0 = E_v\sg_3^z/2 + \hat{W}$ and the gain and decay rates obey $\gamma^+/\gamma^- = e^\chi$. It is convenient to parametrise the effective dissipation rate by the parameter $p = \gamma^+ + \gamma^-$, as in Sec.~\ref{sec:2QE}. Explicit expressions for $\gamma^\pm$ can be found in Appendix~\ref{app:3QE} assuming an underlying reset thermalisation model. 

\subsection{Thermodynamics of precision for the three-qubit engine}
\label{sec:3QE_TUR}

The dynamics under Eq.~\eqref{3QE_master_equation} can be solved analytically to find the power, fluctuations and entropy production in the NESS. As shown in Appendix~\ref{app:3QE}, the results are given by
\begin{align}
    \label{power_3QE}
    \dot{W} & = \Gamma_3 E_v \langle \sg_3^z\rangle_{\rm eq}, \\
    \label{ent_3QE}
    \dot{\Sigma} &= \Gamma_3 \chi \langle \sg_3^z\rangle_{\rm eq}, \\
    \label{fluct_3QE}
    \dot{\Delta}_W & = \Gamma_3 E_v^2 \left[ 1 - \frac{6\Gamma_3 p}{p^2 + 8g^2}\langle \sg_3^z\rangle_{\rm eq}^2\right],
\end{align}
where $\Gamma_3 = 4g^2p/(p^2+8g^2)$ is the characteristic energy transfer rate for the three-qubit machine. As for the two-qubit engine, the proportionality between power and heat currents implies the ideal efficiency $\eta = 1-E_1/E_2$.

Eqs.~\eqref{power_3QE}--\eqref{fluct_3QE} are now easily combined with Eq.~\eqref{bias_qubit3} to obtain
\begin{equation}
    \label{TUR_ratio_3QE}
    \frac{\dot{\Delta}_W}{\dot{W}^2}\dot{\Sigma} = \chi \coth(\chi/2)\left[1 - \frac{6\Gamma_3 p}{p^2 + 8g^2}\tanh^2(\chi/2)\right] \geq 1.245\ldots,
\end{equation}
where the lower bound is derived in Appendix~\ref{app:bounds}. This inequality allows the TUR ratio to take values significantly lower than the classical Markovian result (cf.~Eq.~\eqref{TUR_ratio_2QE}). As we show in Fig.~\ref{fig:TUR_ratio}, such values are achievable: a judicious choice of the coupling ratio $g/p$ leads to substantially reduced power fluctuations relative to the classical TUR. The significance of this finding can be appreciated by rewriting Ineq.~\eqref{TUR_ratio_3QE} in terms of power and efficiency, as in Ineq.~\eqref{efficiency_power_tradeoff}. This shows that, for a given average engine performance, the power fluctuations can at best be reduced to approximately $1.245/2 \approx 62\%$ of those produced by a machine governed by the classical TUR. As in Eq.~\eqref{TUR_ratio_2QE}, the bound in Eq.~\eqref{TUR_ratio_3QE} is saturated only as $\chi\to 0$. We note that the quantum steady-state TUR derived in Ref.~\cite{Guarnieri2019} is always satisfied by Eq.~\eqref{TUR_ratio_3QE}.

The above result demonstrates that adding an extra stage to the transport pathway from engine to load can boost quantum coherence and thereby reduce fluctuations in the machine's output. Similar effects have been reported in the context of thermoelectric devices~\cite{Ptaszynski2018,Agarwalla2018} and autonomous quantum clocks~\cite{Schwarzhans2020}. In the present model, these advantages come at the cost of reduced power output compared to the two-qubit autonomous engine because the optimal coupling regime is $g\lesssim p$, which by assumption is much smaller than the rate of thermalisation (see Appendix~\ref{app:3QE}). However, we stress that this non-essential approximation was made only to simplify the analytical treatment; it is possible that quantum-enhanced reliability can be obtained in strong-coupling regimes of high power, although we leave this question for future work.

\section{Single-spin Otto engine with a harmonic oscillator flywheel}
\label{sec:flywheel}

\subsection{Spin-oscillator engine cycle}

As our final example, we investigate the thermodynamics of precision for a four-stroke heat engine comprising a single qubit working medium coupled to a harmonic oscillator load. Our model is inspired by the experiment reported in Ref.~\cite{Lindenfels2019}, where the qubit comprises two Zeeman spin levels of a trapped ion and the oscillator corresponds to the ion's centre-of-mass degree of freedom. The qubit undergoes an Otto cycle driven by the harmonic motion of the oscillator, which therefore acts as a quantised flywheel. Closely related engine models have also been discussed in Refs.~\cite{Tonner2005,GelbwaserKlimovsky2013}.

\begin{figure}
    \centering
    \includegraphics[width=\linewidth]{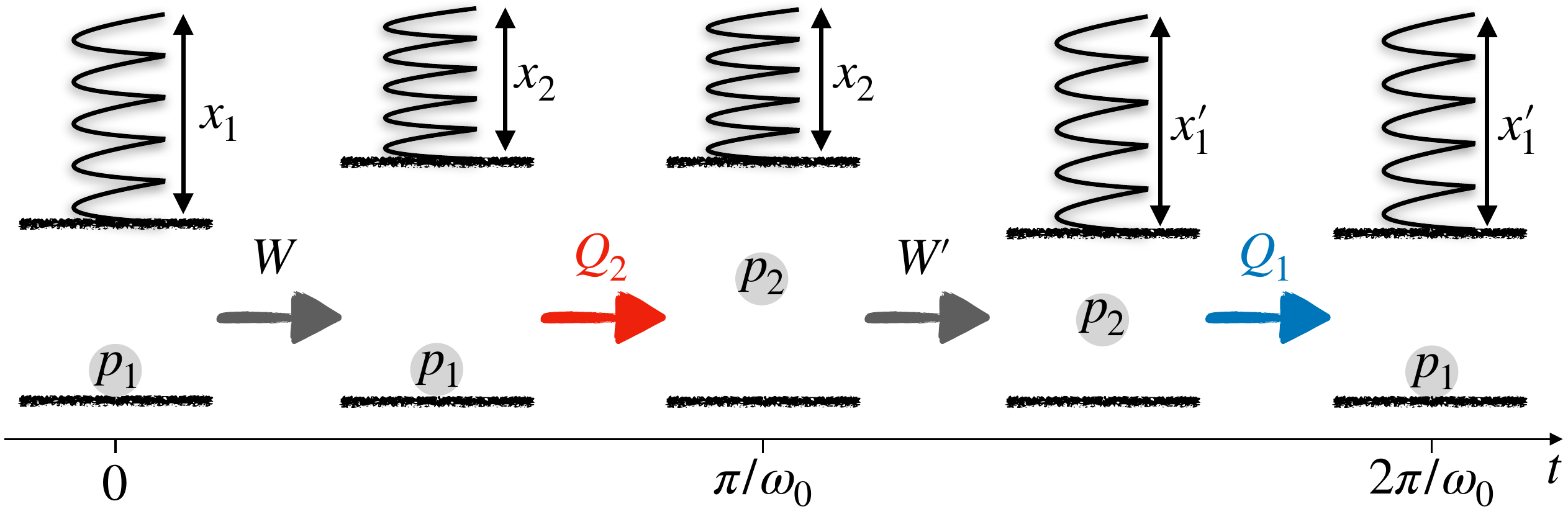}
    \caption{Schematic depiction of an Otto cycle where a qubit working medium is driven by a harmonic oscillator. The harmonic motion of the oscillator, stylised here as a coiled spring of extension $x$, modulates the level spacing of the qubit periodically. The qubit couples to hot and cold baths every half-period, thus changing the excited-state occupation probability (grey circles). The difference in heat absorbed from the hot and cold baths is converted into work done on the spring, driving oscillations of increasing amplitude.}
    \label{fig:flywheel}
\end{figure}

We consider a simple linear coupling between the qubit and oscillator, described by the Hamiltonian
\begin{equation}
\label{spin_flywheel_Hamiltonian}
    \hat{H} = \frac{\omega_z}{2}\sg_z + \omega_0 \adag \a  + \frac{1}{2}\omega_0 d \sg_z \left(\a + \adag\right),
\end{equation}
where $\sg_z$ is a Pauli operator describing the qubit, $\a$ and $\adag$ are canonical ladder operators for the harmonic oscillator, and $d$ is a dimensionless parameter that quantifies the qubit-oscillator coupling strength. A semi-classical depiction of the engine cycle is shown in Fig.~\ref{fig:flywheel}. The qubit is coupled alternately to cold and hot baths at temperatures $T_1$ and $T_2$ every half period, $\delta t/2 = \pi/\omega_0$. In between these isochoric strokes, the qubit is decoupled from the baths and the system evolves freely under the Hamiltonian~\eqref{spin_flywheel_Hamiltonian}. 

In a mean-field picture, the qubit exerts an effective force on the oscillator proportional to $\langle\sg_z\rangle$. Coupling to the heat baths alters the qubit's populations every half-cycle, so that the force oscillates resonantly with the motion. This drives increasingly large oscillations of the flywheel, corresponding to the work output of the engine. In turn, the harmonic motion of the oscillator modulates the effective energy splitting of the qubit by an amount proportional to the displacement $\langle\a+\adag\rangle$.

We focus on the regime of large qubit frequency ($\omega_z\gg \omega_0$), weak qubit-oscillator coupling ($d\ll 1$), and correspondingly small oscillator displacements. We also assume for simplicity that the isochores have a negligible duration and result in perfect thermalisation of the qubit. The heat absorbed during the cold and hot isochores is thus given approximately by the change in the qubit's mean energy, i.e.
\begin{align}
\label{flywheel_heat_1}
    Q_1 & = \left(p_1-p_2\right)\left(\omega_z - \omega_0 d \langle \a+\adag\rangle\right), \\
 \label{flywheel_heat_2}
   Q_2 & = \left (p_2-p_1\right)\left(\omega_z + \omega_0 d \langle \a+\adag\rangle\right),
\end{align}
where $p_j = e^{-\beta_j\omega_z/2}/\mathcal{Z}_j$ is the qubit's excited-state population at temperature $T_j = 1/\beta_j$, while $\langle \a+\adag\rangle$ is the maximum displacement amplitude. In Eqs.~\eqref{flywheel_heat_1} and \eqref{flywheel_heat_2}, we have used a mean-field approximation for the qubit energy and neglected small corrections to the Boltzmann factors due to the oscillator displacement. Hence, the mean energy transferred to the flywheel per cycle is approximately 
\begin{equation}
    \label{work_flywheel_cycle}
    W_{\rm cyc} = Q_1 + Q_2 = 2\omega_0 d(p_2-p_1)\langle \a+\adag\rangle.
\end{equation}
For the purposes of the present study, we consider this energy to be the engine's useful work output. We note that, since the displacement increases on each cycle, the engine's power output increases over time and the cycle is not closed. 

\subsection{Random-walk model of the flywheel dynamics}

Following Ref.~\cite{Lindenfels2019}, we now show that the dynamics of the flywheel can be modelled as a random walk in phase space, under the assumption of fast qubit thermalisation during the isochores. Considering the density matrix of the composite system, the effect of thermalisation is described by the map 
\begin{equation}
    \label{thermalisation_map}
    \mathcal{K}_j\hat{\rho} = \left[ p_j \hat{\Pi}_+ + (1-p_j) \hat{\Pi}_-\right] \hat{\rho}_f ,
\end{equation}
where $j=1,2$ specifies the cold or hot bath, $p_j = e^{-\beta_j\omega_z/2}/\mathcal{Z}_j$ is the corresponding Boltzmann factor, $\hat{\Pi}_{\pm}$ denote projectors onto the ground and excited states of the qubit such that $\sg_z\hat{\Pi}_\pm = \pm\hat{\Pi}_\pm$, and $\hat{\rho}_f = \Tr_{q}[\hat{\rho}]$ is the reduced state of the flywheel obtained by tracing over the qubit. Between the isochores, the state evolves freely under the Hamiltonian~\eqref{spin_flywheel_Hamiltonian} over a time interval $\delta t/2$. This generates the unitary map $\mathcal{U}\hat{\rho} = e^{-i \hat{H} \delta t/2} \hat{\rho}e^{i \hat{H} \delta t/2}$, with
\begin{equation}
    \label{flywheel_unitary}
    e^{-i \hat{H} \delta t/2} = \hat{\Pi}_+ \hat{P} \hat{D}(+d) + \hat{\Pi}_- \hat{P} \hat{D}(-d),
\end{equation}
where $\hat{D}(\pm d) = e^{\pm d(\adag-\a)}$ and $\hat{P} = e^{i\pi \adag\a}$. Physically, $\hat{D}(\pm d)$ enacts a displacement of the oscillator in phase space, while the parity operator $\hat{P}$ reverses the direction of motion every half-period.

\begin{figure}
    \centering
    \includegraphics[width=0.8\linewidth]{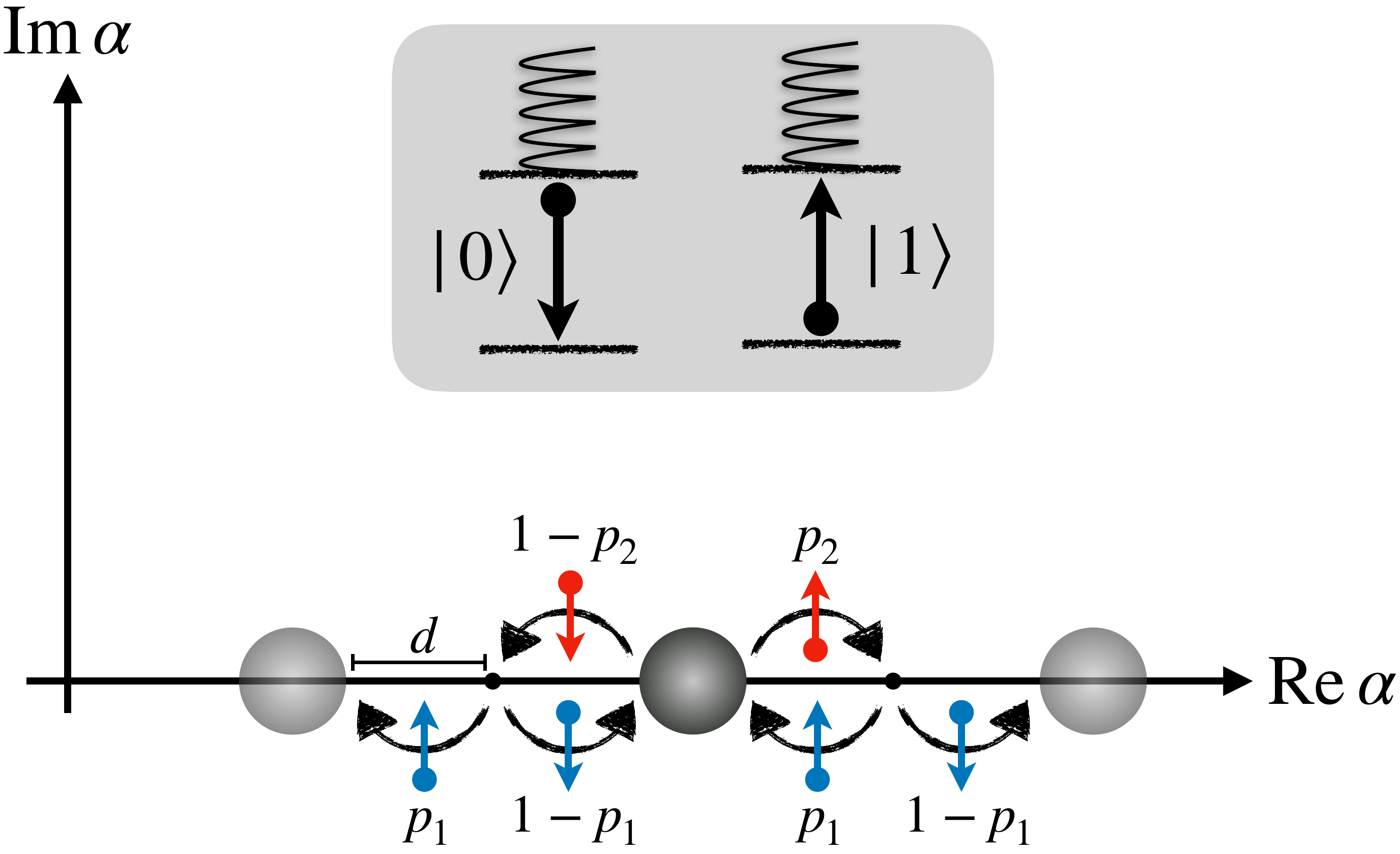}
    \caption{The flywheel dynamics can be modelled as a discrete-time random walk in phase space, where each value of $\alpha$ represents a coherent state of the oscillator. The state of the qubit can be visualised as a spin pointing up or down, whose orientation determines the direction of the force on the oscillator (inset). If the spin flips its orientation after half the cycle, the effective driving force is resonant with the oscillatory motion, thus changing the displacement in phase space by $\pm 2d$. If the spin's orientation does not change, the effective force is constant and the displacement is unaltered after a full cycle.}
    \label{fig:random_walk}
\end{figure}

The evolution over one full period is found by concatenating the thermalisation and unitary maps in turn, yielding $\hat{\rho}^{(N)} = \mathcal{U} \mathcal{K}_1 \mathcal{U} \mathcal{K}_2 \hat{\rho}^{(N-1)}$ as the state after $N$ engine cycles (i.e.~$N$ oscillation periods). Tracing over the qubit, we obtain a recursion relation for the flywheel state
\begin{align}
    \label{flywheel_map}
    \hat{\rho}_f^{(N)} & = p_0 \hat{\rho}_f^{(N-1)} + p_+\hat{D}(+2d)\hat{\rho}_f^{(N-1)}\hat{D}^\dagger(+2d) \notag \\ & 
    \quad + p_-\hat{D}(-2d)\hat{\rho}_f^{(N-1)}\hat{D}^\dagger(-2d).
\end{align}
This describes a discrete-time random walk in phase space, where $p_+ = p_2(1-p_1)$ is the probability of taking a forward step, $p_- = p_1(1-p_2)$ is the probability of a backward step, while with probability $p_0 = 1-p_+-p_-$ the state does not change. As depicted in Fig.~\ref{fig:random_walk}, these probabilities have a natural interpretation in terms of ``spin-flip'' processes during the isochores. Each forward step of size $2d$ corresponds to an increase of the oscillation amplitude, whereby the qubit working medium deposits energy into the flywheel. The ratio of the forward and backward rates is given by $p_+/p_- = e^{\chi}$ where, by analogy with Eq.~\eqref{positive_entropy_production}, we define the bias parameter
\begin{equation}
    \label{chi_flyweel}
    \chi = (\beta_1 - \beta_2)\omega_z\geq 0.
\end{equation}

Assuming the oscillator is initialised in its ground state, after $N$ engine cycles it is in a coherent state $\ket{\alpha_N}$, where $\alpha_N$ is a real random variable representing the distance travelled after $N$ steps of the random walk. Normal-ordered quantum expectation values may thus be computed as $\langle (\adag)^p\a^q\rangle = \mathbb{E}[\alpha_N^{q+p}]$, where $\mathbb{E}[\bullet]$ represents an average over random-walk trajectories. Such averages can be found systematically by taking derivatives of the moment generating function $G_N(s) = \mathbb{E}[e^{ s\alpha_N}]$ at $s=0$. Since each step of the random walk is independent and identically distributed, we can write $G_N(s) = [G_1(s)]^N$, where the generating function for one cycle is
\begin{equation}
    \label{char_function}
    G_1(s) = 1+2\sinh(ds)\left(p_+ e^{ds} - p_- e^{-ds}\right).
\end{equation}

\subsection{Thermodynamics of precision for the random walk}

The stochastic nature of the flywheel dynamics implies the existence of fluctuations in the engine's work output, which we now relate to the entropy production. The total energy change of the flywheel after $N$ cycles is $W_N = \omega_0 \langle \hat{n}\rangle$, where $\hat{n} = \adag\a$, which can be found using Eq.~\eqref{char_function}. For large $N$, we obtain to leading order
\begin{equation}
    \label{work_flywheel}
        W_N = 4\omega_0 d^2(p_+-p_-)^2 N^2 + \mathcal{O}(N).
\end{equation}
This quadratic increase in energy with $N$ follows from a linear growth of displacement, i.e.~$\langle \hat{a}\rangle = 2 d(p_+-p_-)N$, as expected for a biased random walk over coherent states. We note that the change in energy over a single cycle is therefore $W_{\rm cyc} = W_{N+1} - W_N \approx 4\omega_0d(p_+ - p_-)\langle \a\rangle$, in exact agreement with Eq.~\eqref{work_flywheel_cycle}. The fluctuations of the flywheel's energy are given by $\Delta_{W_N} = \omega_0^2 \langle \hat{n}^2\rangle - W_N^2$. Explicitly, we find that
\begin{align}
    \label{fluc_flywheel}
    \Delta_{W_N} & = 64 \omega_0^2 d^4 (p_+-p_-)^2\left[p_++p_- - (p_+-p_-)^2\right]N^3 \notag \\ 
    & \quad + \mathcal{O}(N^2).
\end{align}
The entropy production after $N$ cycles is given by $\Sigma_N = -N(\beta_1 Q_1+\beta_2Q_2)$. Using Eqs.~\eqref{flywheel_heat_1} and \eqref{flywheel_heat_2}, this can be approximated to leading order by
\begin{equation}
 \label{flywheel_entropy_prod}
 \Sigma_N \approx (\beta_1 - \beta_2)(p_+-p_-)\omega_z N,
\end{equation}
since the bare Zeeman energy $\omega_z$ is assumed to dominate the contribution from the oscillator displacement. 

To obtain an expression connecting power fluctuations and entropy production rate, we convert Eqs.~\eqref{work_flywheel}--\eqref{flywheel_entropy_prod} into coarse-grained rates of change over an engine cycle, e.g. $\dot{W} = (W_{N+1}-W_N)/\delta t$, with $\delta t=2\pi/\omega_0$. We thus obtain the TUR ratio for the flywheel at leading order for large $N$:
\begin{equation}
    \label{flywheel_TUR}
    \frac{\dot{\Delta}_{W}}{\dot{W}^2} \dot{\Sigma} = 3\chi \left[ \coth(\chi/2) - \left(p_+ - p_-\right) \right] + \mathcal{O}(N^{-1}).
\end{equation}
This differs in two important respects from the analogous expression $\chi\coth(\chi/2)$, which holds for the classical, continuous-time random walk~\cite{Barato2015}. The second term inside the square brackets of Eq.~\eqref{flywheel_TUR}, which reduces the relative fluctuations, appears due to the discrete-time nature of the evolution under consideration. This term vanishes in the limit of a continuous-time process where $p_\pm \sim \dd t\to 0$. This illustrates the fact that a continuous-time random walk exhibits enhanced fluctuations because of the random timing of the steps~\cite{Chiuchiu2018}. 

\begin{figure}
	\centering
	\includegraphics[width=\linewidth]{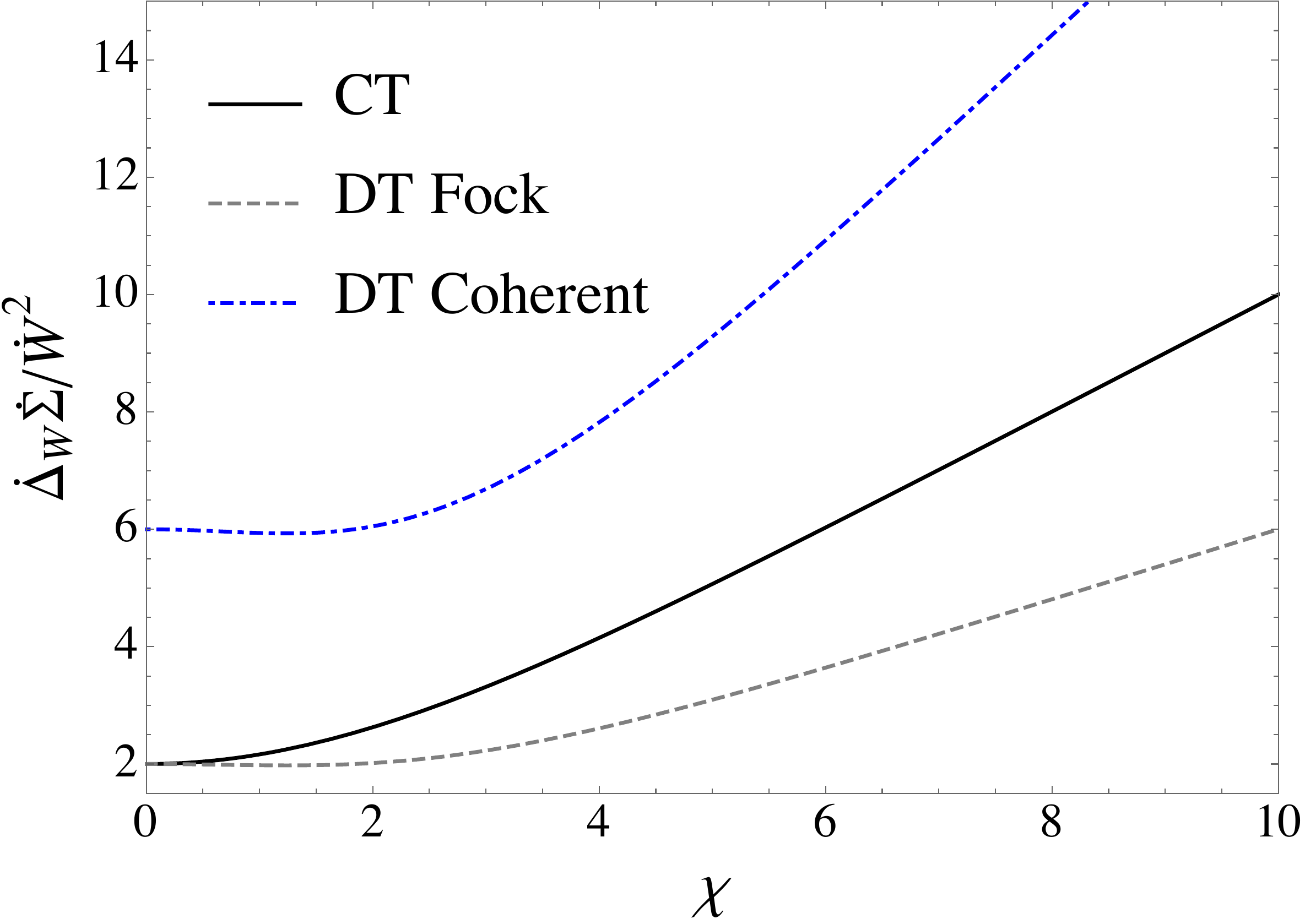}
	\caption{Thermodynamics of precision for random walks. The TUR ratio after a large number of steps $N$ is plotted against the bias parameter $\chi$, which quantifies the entropy production per step. The result for the flywheel (blue dot-dashed line), corresponding to a discrete-time (DT) random walk over coherent states, is compared to a DT random walk over Fock states (grey dashed line) and a classical continuous-time (CT) random walk (solid black line). The parameters are chosen so that $p_0 = 0.6$. \label{fig:TUR_flywheel}}
\end{figure}

The other important feature in Eq.~\eqref{flywheel_TUR} is the overall factor of three multiplying the expression. The physical origin of this factor is the fact that the random walk takes place over coherent states in phase space, rather than energy eigenstates in Fock space. Even a pure coherent state $\ket{\alpha}$ has a finite energy uncertainty given by $\bra{\alpha}\hat{n}^2\ket{\alpha}-\bra{\alpha}\hat{n}\ket{\alpha}^2 = |\alpha|^2$. The growth of energetic fluctuations is thus faster than would be obtained for a random walk between Fock states. This is easily shown by interpreting the generating function~\eqref{char_function} as describing a single step in Fock space, $\ket{n}\to\ket{n\pm 1}$, which leads to an expression precisely three times smaller than Eq.~\eqref{flywheel_TUR}. The TUR ratio for the flywheel is compared to a random walk over Fock states in Fig.~\ref{fig:TUR_flywheel}.

\section{Discussion}

When considering heat engines at the smallest scales, the fluctuations in output power may be as important as the average performance. Here we have derived explicit expressions for these fluctuations, relating them to the average rate of entropy production for several quantum heat engine models of current interest. Our results exemplify the entropic penalty associated with a reliable power output. In particular, just as for many classical heat engines~\cite{Shiraishi2016, Holubec2017,Pietzonka2018,Holubec2018,Miller2020,Abiuso2020}, approaching the Carnot efficiency (i.e.~zero entropy production) at finite power comes at the cost of catastrophically large fluctuations. This conclusion holds for both autonomous (Secs.~\ref{sec:2QE} and~\ref{sec:3QE}) and cyclic (Sec.~\ref{sec:flywheel}) few-qubit engines.

Interestingly, however, quantum mechanics opens the possibility of reducing an engine's relative power fluctuations below the level allowed by classical stochastic thermodynamics. In particular, Eq.~\eqref{TUR_ratio_coherence} clarifies two different ways in which autonomous few-qubit machines can surpass the classical TUR. Either one can reduce the occupation $\langle\hat{N}\rangle$ of the virtual qubit, or one can boost the coherent current $\langle \hat{C}\rangle$. This conclusion should apply quite generally to more complicated multi-level machines, which can also be analysed in terms of virtual qubits~\cite{Brunner2012,Silva2016,Usui2020}. We demonstrated this principle explicitly by considering two variants of the two-qubit engine in Secs.~\ref{sec:2QE_local_lindblad} and~\ref{sec:3QE}. Both of these exhibit smaller power fluctuations because the effect of decoherence is reduced, either as a consequence of the thermalisation model (Sec.~\ref{sec:2QE_local_lindblad}) or the geometry of the engine itself (Sec.~\ref{sec:3QE}). This suggests a general design principle for reducing fluctuations in autonomous thermal machines by boosting coherent transport, e.g.~via the introduction of additional stages in the transport pathway between engine and load~\cite{Ptaszynski2018,Schwarzhans2020}. According to Eq.~\eqref{TUR_ratio_coherence}, it may also be possible to tame power fluctuations by designing the engine so that the virtual qubit's occupation is reduced, yet we have not found explicit models where this occurs.

Nevertheless, quantum coherence can lead to drawbacks as well as advantages. We showed in Sec.~\ref{sec:flywheel} that an engine which drives transitions between coherent states of the load suffers from greater fluctuations, as compared to a machine whose load remains incoherent in its local energy eigenbasis. This is simply because coherent superpositions of energy eigenstates have an intrinsic energy uncertainty above and beyond the thermodynamic uncertainty introduced by coupling to heat reservoirs. We note, however, that such coherences do have some potential to be extracted as useful work with an appropriate protocol~\cite{Francica2020}. A careful consideration of this problem~\cite{Niedenzu2019} would lead us to evaluate the ergotropy~\cite{Allahverdyan2004} or the non-equilibrium free energy~\cite{Esposito2011} of the load, as opposed to its mere energy statistics, a question that we leave for future work.  

From a more foundational perspective, our findings highlight the importance of fluctuations in certifying the non-classical behaviour of quantum thermal machines~\cite{Uzdin2015,Smirne2018,Klatzow2019,Verteletsky2020,Lostaglio2020,Levy2020}. In contrast to an engine's average performance~\cite{Gonzalez2019}, power fluctuations that violate a classical TUR cannot be emulated by any classical (Markovian) stochastic model and thus constitute an unambiguous quantum advantage. We note that the use of fluctuations to tease out the quantum character of dynamical processes has a distinguished history, e.g.~in the field of quantum optics~\cite{HanburyBrown1956,*Twiss1957,Kimble1977}. We therefore hope that our work will stimulate further research on the general characterisation and effective suppression of fluctuations in quantum thermal machines.

\begin{acknowledgments}
We are grateful to Stephen R.~Clark and Paul Skrzypczyk for enlightening discussions. We acknowledge funding from the European Research Council Starting Grant ODYSSEY (G.~A.~758403). JG is supported by a SFI-Royal Society University Research Fellowship.
\end{acknowledgments}

\appendix

\section{ Solution for the two-qubit engine}
\label{app:2QE}

\subsection{Reset model}
\label{app:reset_model}

For completeness, in this Appendix we detail the solution of the two-qubit engine modelled by the reset master equation. The results are equivalent to those already obtained by Brunner et al.~\cite{Brunner2012}, but it is convenient to recast them here in our notation. Starting from the master equation [Eq.~\eqref{reset_master_equation_2QE}] written as $\dd\rrho/\dd t = -i[\H,\rrho] + \sum_{j=1,2} \mathcal{D}_j\rrho$, the dynamics of any operator $\hat{O}$ in the Heisenberg picture is determined by the equation of motion ${\dd\hat{O}/\dd t = i[\H,\hat{O}] + \sum_{j}\mathcal{D}_j^\dagger\hat{O}}$, where the adjoint dissipator is defined implicitly by $\Tr[\hat{O}\mathcal{D}_j\rrho] = \Tr[\rrho\mathcal{D}_j^\dagger \hat{O}]$; see Ref.~\cite{BreuerPetruccione} for details. 

The Heisenberg equation for the load Hamiltonian is simply $\dd \hat{W}/\dd t = gE_v\hat{C}$, where the current operator is defined in Eq.~\eqref{current_operator}. The state of the engine qubits is determined by the bias [Eq.~\eqref{bias_virtual}] and occupation [Eq.~\eqref{occupation_virtual}] of the virtual qubit, as well as the total spin $\hat{S} = \tfrac{1}{2}(\sg_1^z + \sg_2^z)$ (equivalent to the bias of an ``anti-virtual qubit''~\cite{Brunner2012}). Together with the current, these quantities obey the coupled equations\begin{align}
    \label{current_EOM}
    \frac{\dd \hat{C}}{\dd t} & = 2g \hat{Z} - 2p \hat{C}, \\
    \label{bias_EOM}
    \frac{\dd \hat{Z}}{\dd t} & = p\left( \langle \hat{Z}\rangle_{\rm eq} - \hat{Z}\right) - 2g\hat{C},\\
    \label{occupation_EOM}
    \frac{\dd \hat{N}}{\dd t} & = p\left(1- 2\hat{N} +\langle \hat{Z}\rangle_{\rm eq} \hat{Z} - \langle \hat{S}\rangle_{\rm eq} \hat{S} \right),\\
    \label{antivirtual_EOM}
    \frac{\dd \hat{S}}{\dd t} & = p\left(\langle \hat{S}\rangle_{\rm eq}  - \hat{S}\right),
\end{align}
where the equilibrium averages denoted by, for example, $\langle \hat{Z}\rangle_{\rm eq} = \Tr(\hat{Z}\tau_1\otimes\tau_2)$, can be evaluated using the formula $\langle \sg_j^z\rangle_{\rm eq} = -\tanh(E_j/2T_j)$. The quasi-stationary state is found by setting the mean value of the above derivatives above to zero, yielding the solutions $\langle \hat{S}\rangle = \langle \hat{S}\rangle_{\rm eq}$ and
\begin{align}
    \label{mean_current}
    \langle \hat{C}\rangle & = \frac{gp}{p^2+2g^2}\langle \hat{Z}\rangle_{\rm eq},\\
    \label{mean_bias}
    \langle \hat{Z}\rangle & = \frac{p^2}{p^2+2g^2}\langle \hat{Z}\rangle_{\rm eq},\\
    \label{mean_occupation}
    \langle \hat{N}\rangle & = \langle \hat{N}\rangle_{\rm eq} - \frac{g^2}{p^2+2g^2} \langle \hat{Z}\rangle_{\rm eq}^2.
\end{align}
The mean rate of energy transfer to the load is then given by $\dot{W} = \Gamma_2 E_v\langle Z\rangle_{\rm eq}$, with $\Gamma_2 = g^2p/(p^2+2g^2)$, which is equivalent to Eq.~\eqref{work_2QE}.

To find the fluctuations, we consider the equation of motion for $\hat{W}^2$, which is $\dd \hat{W}^2/\dd t = g E_v \hat{K}$, with $\hat{K} = \{\hat{W},\hat{C}\}$. We also define the operator $\hat{\Omega} = \hat{Z}\hat{W}$ in the Schr\"odinger picture, which in the Heisenberg picture is coupled to $\hat{K}$ via the equations
\begin{align}
    \label{K_EOM}
    \frac{\dd \hat{K}}{\dd t} & = 2g\left(2\hat{\Omega} + E_v \hat{N}\right) - 2p\hat{K},\\
    \label{Omega_EOM}
    \frac{\dd \hat{\Omega}}{\dd t} & = p\left ( \langle \hat{Z}\rangle_{\rm eq} \hat{W} - \hat{\Omega} \right) - g\hat{K}.
\end{align}
After eliminating $\hat{\Omega}$ from the equations and focussing on asymptotically long times where the solutions in Eqs.~\eqref{mean_current}--\eqref{mean_occupation} hold, we obtain
\begin{align}
\label{K_SHO_EOM}
    \left[\frac{\dd^2}{\dd t^2}  + 3p\frac{\dd}{\dd t}  + 2p^2+4g^2 \right]\langle \hat{K}\rangle = 2g p E_v \left( 2\Gamma_2 t\langle \hat{Z}\rangle_{\rm eq}^2 + \langle \hat{N}\rangle \right).
\end{align}
This describes a damped harmonic oscillator under a driving force given by the right-hand side of the equation. Considering times long enough for transient oscillations to decay to zero, i.e.\ $pt\gg 1$, the solution is easily found by using the ansatz $\langle \hat{K}\rangle = a + bt$ and solving for the constants $a$ and $b$. This procedure ultimately yields
\begin{equation}
    \label{fluctuations_solution}
    \frac{\dd}{\dd t} \langle W^2\rangle  = \frac{\dd}{\dd t}\langle \hat{W}\rangle^2 + \Gamma_2 E_v^2 \left[\langle \hat{N}\rangle_{\rm eq} - \frac{2g^2(g^2+2p^2)}{(p^2+2g^2)^2}\langle \hat{Z}\rangle_{\rm eq}^2 \right],
\end{equation}
which is equivalent to Eq.~\eqref{fluct_2QE}.

Finally, we need the energy currents entering the system from each bath, $j=1,2$, defined by $\dot{Q}_j = \langle \mathcal{D}_j^\dagger\hat{H}\rangle$. A simple calculation reveals that
\begin{equation}
    \label{energy_current_general}
    \dot{Q}_j = p E_j \left( \langle \sg_j^z\rangle_{\rm eq} - \langle \sg_j^z\rangle \right) - p\langle \H_{\rm int}\rangle.
\end{equation}
The Heisenberg equation for $\H_{\rm int}$ is $\dd \H_{\rm int}/\dd t = -2p \hat{H}_{\rm int}$, whose stationary solution is $\langle \H_{\rm int}\rangle = 0$. Eq.~\eqref{energy_current_general} therefore recovers the local energy current defined by Brunner et al.~\cite{Brunner2012}, i.e.~$\dot{Q}_j = \langle \mathcal{D}_j^\dagger\hat{H}_0\rangle$. (Note that the mean interaction energy vanishes only because of the assumption of resonant interactions, i.e.\ $[\H_0,\H_{\rm int}] = 0$, which in the language of collisional models means that the qubit resets do not perform work on average~\cite{Barra2015, Barra2018, DeChiara2018}.) It is then straightforward to verify from the Heisenberg equations for $\langle \sg_j^z\rangle$ that the stationary heat currents are 
\begin{align}
    \label{J_solution}
    \dot{Q}_1 = -gE_1 \langle \hat{C}\rangle,\quad \dot{Q}_2 = gE_2\langle \hat{C}\rangle.
\end{align}

\subsection{Local Lindblad equation}
\label{app:local_lindblad}

We now carry out the same calculation for the local Lindblad dissipation model defined by Eq.~\eqref{local_ME_Lindblad_2QE}, which differs from the reset model of Eq.~\eqref{reset_master_equation_2QE} by the absence of the dephasing terms proportional to $\gamma^z$. These terms affect only off-diagonal operators in the computational basis. Therefore, only the equations of motion for $\hat{C}$ and $\hat{K}$ are altered, being given by
\begin{align}
    \label{C_EOM_local_lind}
    \frac{\dd \hat{C}}{\dd t} & = 2g \hat{Z} - p \hat{C}, \\
    \label{K_EOM_local_lind}
    \frac{\dd \hat{K}}{\dd t} & = 2g\left(2\hat{\Omega} + E_v \hat{N}\right) - p\hat{K}.
\end{align}
We see that the effective decoherence rate is cut in half. All other equations of motion and manipulations proceed as in the previous section. We find the solutions
\begin{align}
    \label{mean_current_ll}
    \langle \hat{C}\rangle & = \frac{2gp}{p^2+4g^2}\langle \hat{Z}\rangle_{\rm eq},\\
    \label{mean_bias_ll}
    \langle \hat{Z}\rangle & = \frac{p^2}{p^2+4g^2}\langle \hat{Z}\rangle_{\rm eq},\\
    \label{mean_occupation_ll}
    \langle \hat{N}\rangle & = \langle \hat{N}\rangle_{\rm eq} - \frac{2g^2}{p^2+4g^2} \langle \hat{Z}\rangle_{\rm eq}^2,
\end{align}
from which we obtain the power $\dot{W} = gE_v\langle \hat{C}\rangle = \Gamma_2' E_v \langle \hat{Z}\rangle_{\rm eq}$, with $\Gamma_2' = 2g^2p/(p^2+4g^2)$. The heat currents are given by Eq.~\eqref{J_solution} together with Eq.~\eqref{mean_current_ll}.

The above solutions are then used to find an effective equation of motion for $\langle \hat{K}\rangle$, which reads as
\begin{align}
\label{K_SHO_EOM_ll}
    \left[\frac{\dd^2}{\dd t^2}  + 2p\frac{\dd}{\dd t}  + p^2+4g^2 \right]\langle \hat{K}\rangle = 2g p E_v \left( 2\Gamma_2' t\langle \hat{Z}\rangle_{\rm eq}^2 + \langle \hat{N}\rangle \right).
\end{align}
Postulating an asymptotic solution of the form $\langle \hat{K}\rangle = a+bt$, we eventually find
\begin{equation}
    \label{Delta_W_dot_LL}
    \dot{\Delta}_W = \Gamma_2'E_v^2 \left[\langle \hat{N}\rangle_{\rm eq} - \frac{2g^2(5p^2+4g^2)}{(p^2+4g^2)^2}\langle  \hat{Z}\rangle_{\rm eq}^2 \right],
\end{equation}
which is equivalent to Eq.~\eqref{fluct_LL}.

\section{Bounds for the TUR ratio}
\label{app:bounds}

In this Appendix, we derive bounds for the TUR ratios given in the main text. Let us begin with the two-qubit engine described by a reset thermalisation model in Eq.~\eqref{reset_master_equation_2QE}, whose TUR ratio is given in Eq.~\eqref{TUR_ratio_2QE}. We first note from Eqs.~\eqref{bias_mean} and \eqref{occupation_mean} that $\langle \hat{Z}\rangle_{\rm eq} \geq \tfrac{1}{2}\tanh(\chi/2)$, where equality is reached only in the limit of vanishing bias and work output, i.e. $\beta_jE_j\to 0$. However, this simple inequality allows us to bound the TUR ratio as
\begin{equation}
\label{TUR_ratio_bound2QE}    
    \frac{\dot{\Delta}_W}{\dot{W}^2}\dot{\Sigma} \geq \chi \coth(\chi/2) \left[ 1 - \frac{1}{2}f(g/p)\tanh^2(\chi/2) \right],
\end{equation}
where we defined the function
\begin{equation}
    \label{bounding_function}
    f(r) = \frac{2r^2(2+r^2)}{(1+2r^2)^2}.
\end{equation}
The above function is positive and has a maximum equal to $\max_r f(r) = 2/3$, which is attained at $r=1$. It follows that 
\begin{equation}
\label{TUR_ratio_bound2QE_sol}    
    \frac{\dot{\Delta}_W}{\dot{W}^2}\dot{\Sigma} \geq \chi \coth(\chi/2) \left[ 1 - \frac{1}{3}\tanh^2(\chi/2) \right] \geq 2,
\end{equation}
where the final equality is saturated only in the limit $\chi\to 0$. 

For the local Lindblad equation~\eqref{local_ME_Lindblad_2QE}, we can write an expression analogous to Ineq.~\eqref{TUR_ratio_bound2QE} but with a different function
\begin{equation}
    f(r) = \frac{2r^2(5+4r^2)}{(1+4r^2)^2},
\end{equation}
whose maximum value $\max_r f(r) = 25/32$ is attained at $r = \sqrt{5/12}$. This leads to the bound
\begin{equation}
\label{TUR_ratio_bound2QE_LL}    
    \frac{\dot{\Delta}_W}{\dot{W}^2}\dot{\Sigma} \geq \chi \coth(\chi/2) \left[ 1 - \frac{25}{64}\tanh^2(\chi/2) \right] \geq 1.982\ldots \; ,
\end{equation}
which can be verified numerically. In practice, we find that any violations of the classical bound are very small and occur in a restricted region of the parameter space. 

Finally, we move to the three-qubit engine model of Sec.~\ref{sec:3QE}. Here, we have an exact equality given by Eq.~\eqref{TUR_ratio_3QE}, which is of the form of the right-hand side of Ineq.~\eqref{TUR_ratio_bound2QE_sol} with
\begin{equation}
    f(r) = \frac{24r^2}{(1+8r^2)^2}.
\end{equation}
This function is upper-bounded by $\max_r f(r) = 3/4$, which occurs when $r=1/2\sqrt{2}$. As a result, we obtain the bound
\begin{equation}
\label{TUR_ratio_bound3QE}    
    \frac{\dot{\Delta}_W}{\dot{W}^2}\dot{\Sigma} \geq \chi \coth(\chi/2) \left[ 1 - \frac{3}{4}\tanh^2(\chi/2) \right] \geq 1.245\ldots \; .
\end{equation}

\section{Solution for the three-qubit engine}
\label{app:3QE}

\subsection{Master equation derivation}
\label{app:3QE_ME}

In this Appendix, we detail the solution of the three-qubit engine model in the limit of fast thermalisation (i.e.~weak coherent coupling). We begin by sketching the derivation of the effective master equation. Let $\hat{R}(t)$ denote the total density matrix of the three qubits and the load. We consider a reset thermalisation model for the qubits, described by a dissipator $\mathcal{D}\hat{R} = \sum_{j=1}^2 p'[\ttau_j \otimes \Tr_j(\hat{R}) - \hat{R}]$ and thermalisation rate $p'$. We also define the Hamiltonian superoperators $\mathcal{H}'_0$, $\mathcal{H}_{\rm int}$ and $\mathcal{V}$ in terms of the commutators with the corresponding Hamiltonians in Eqs.~\eqref{3QE_h0}--\eqref{3QE_Hint}, e.g.~$\mathcal{H}'_0\hat{R} = -i[\H_0',\hat{R}]$. The master equation can then be written in the form $\dd \hat{R}/\dd t = (\mathcal{L}_0+ \mathcal{V} + \mathcal{H}_{\rm int})\hat{R}$, where $\mathcal{L}_0 = \mathcal{H}_0'+\mathcal{D}$. 

The assumption of fast thermalisation means that $\mathcal{L}_0 \gg \mathcal{V},\mathcal{H}_{\rm int}$, so that the damped qubits mostly remain in thermal equilibrium and are only weakly perturbed by the interaction terms. This idea can be formalised using standard projection-operator techniques~\cite{BreuerPetruccione,Rivas2010}. We define a projector by
\begin{equation}
    \label{projector}
    \mathcal{P}\hat{R} = \ttau_{1}\otimes \ttau_2\otimes \Tr_{12}(\hat{R}),
\end{equation}
which satisfies the easily verified properties and relations
\begin{align}
    \label{projector_properties}
    [\mathcal{P},\mathcal{L}_0] = 0,\quad 
    [\mathcal{P},\mathcal{H}_{\rm int}]  = 0, \quad
    \mathcal{P}\mathcal{V}\mathcal{P}  = 0.
\end{align}
We also define the orthogonal projector $\mathcal{Q} = 1-\mathcal{P}$, such that $\mathcal{P}^2 = \mathcal{P}$, $\mathcal{Q}^2 = \mathcal{Q}$, and $\mathcal{Q}\mathcal{P} = \mathcal{P}\mathcal{Q}=0$. Let us move to an interaction picture via the transformation $\tilde{R}(t) = e^{-\mathcal{L}_0 t}\hat{R}(t),$ where the dynamics is described by the master equation
\begin{equation}
    \label{3QE_total_ME}
    \frac{\dd \tilde{R}}{\dd t} = \tilde{\mathcal{V}}(t) \tilde{R}(t) + \tilde{\mathcal{H}}_{\rm int}(t) \tilde{R}(t).
\end{equation}
Above, tildes denote superoperators in the interaction picture, e.g.~$\tilde{\mathcal{V}}(t) = e^{-\mathcal{L}_0t}\mathcal{V} e^{\mathcal{L}_0t}$. Applying projectors to both sides of the master equation, inserting appropriate factors of $1 = \mathcal{P} +\mathcal{Q}$, and using the properties~\eqref{projector_properties}, we find that
\begin{align}
\label{P_EOM}
    \frac{\dd }{\dd t}\mathcal{P}\tilde{R} & = \mathcal{P}\tilde{\mathcal{V}}(t) \mathcal{Q}\tilde{R}(t) + \tilde{\mathcal{H}}_{\rm int}(t) \mathcal{P}\tilde{R}(t) ,\\
    \label{Q_EOM}
    \frac{\dd }{\dd t}\mathcal{Q}\tilde{R} & =\mathcal{Q}\tilde{\mathcal{V}}(t) \mathcal{Q}\tilde{R}(t) +\mathcal{Q}\tilde{\mathcal{V}}(t) \mathcal{P}\tilde{R}(t)  + \tilde{\mathcal{H}}_{\rm int}(t) \mathcal{Q}\tilde{R}(t).
\end{align}
The solution of Eq.~\eqref{Q_EOM} with initial condition $\mathcal{Q}\hat{R}(0)=0$ is
\begin{align}
    \label{Q_Soln}
    & \mathcal{Q}\tilde{R}(t)= \int_0^t \dd t'\, \mathcal{G}(t,t')\tilde{\mathcal{V}}(t') \mathcal{P}\tilde{R}(t'),  \\
    \label{propagator}
    & \mathcal{G}(t,t') = {\rm T}\exp \left[ \int_{t'}^t\dd t'' \mathcal{Q}\left( \tilde{\mathcal{V}}(t'')+\tilde{\mathcal{H}}_{\rm int}(t'')\right)\mathcal{Q}\right],
\end{align}
where we have used Eq.~\eqref{projector_properties} to write $\mathcal{Q}\tilde{\mathcal{V}} \mathcal{P} = (1-\mathcal{P})\tilde{\mathcal{V}} \mathcal{P} = \tilde{\mathcal{V}} \mathcal{P}$. Plugging the above solution back into Eq.~\eqref{P_EOM} yields
\begin{align}
    \label{P_closed}
    \frac{\dd }{\dd t}\mathcal{P}\tilde{R} & = \tilde{\mathcal{H}}_{\rm int}(t) \mathcal{P}\tilde{R}(t) + \int_0^t \dd t'\, \mathcal{P}\tilde{\mathcal{V}}(t) \mathcal{G}(t,t')\tilde{\mathcal{V}}(t') \mathcal{P}\tilde{R}(t'),\notag \\
    & \approx \tilde{\mathcal{H}}_{\rm int}(t) \mathcal{P}\tilde{R}(t) + \int_0^t \dd t' \, \mathcal{P}\tilde{\mathcal{V}}(t)\tilde{\mathcal{V}}(t-t') \mathcal{P}\tilde{R}(t-t'),\notag \\
    & \approx  \tilde{\mathcal{H}}_{\rm int}(t) \mathcal{P}\tilde{R}(t) + \int_0^\infty \dd t' \, \mathcal{P}\tilde{\mathcal{V}}(t)\tilde{\mathcal{V}}(t-t') \mathcal{P}\tilde{R}(t).
\end{align}
On the second line, we have expanded $\mathcal{G}(t,t')$ to lowest order in the small quantities $\mathcal{V}$ and $\mathcal{H}_{\rm int}$, and shifted the integration variable as $t'\to t-t'$. On the third line, we have invoked the Markov approximation by assuming that the integrand decays on a timescale much shorter than the characteristic evolution timescale of $\tilde{R}(t)$. Since the former timescale is given by $(p')^{-1}$ and the latter is determined by the inverse of $k$ and $g$, this approximation is consistent with our starting assumption that $p'\gg g,k$. 

Finally, we trace over qubits 1 and 2 and transform back to the Schr\"odinger picture to obtain the master equation~\eqref{3QE_master_equation}, with $\rrho(t) = \Tr_{12}[\hat{R}(t)]$ and the rates given by
\begin{equation}
    \gamma^\pm = \frac{k^2 e^{\pm\chi/2}}{p' \mathcal{Z}_1 \mathcal{Z}_2} .
\end{equation}
The self-consistency of the above derivation requires that ${\gamma^+ + \gamma^- = p \ll p'}$, hence the energy transfer dynamics described by Eq.~\eqref{3QE_master_equation} is necessarily much slower than the underlying thermalisation processes. We note that other choices for the dissipator $\mathcal{D}$ that obey $\mathcal{D} \ttau_1\otimes\ttau_2 = 0$ would lead to a master equation of the same form, but with different expressions for the rates. 
\subsection{Asymptotic solution}
\label{app:3QE_soln}

The solution proceeds straightforwardly according to the methods of Appendix~\ref{app:2QE}. The mean energy of the load follows from the equation of motion $\dd \hat{W}/\dd t = gE_v\langle \hat{C}\rangle$, with the current operator now given by
\begin{equation}
    \label{current_3QE}
    \hat{C} = ig\left(\sg_3^+\hat{A} - \sg_3^- \hat{A}^\dagger \right).
\end{equation}
The current and qubit bias obey the coupled differential equations
\begin{align}
\label{C_EOM_3}
    \frac{\dd \hat{C}}{\dd t} & = 2 g \sg_3^z - \frac{p}{2}\hat{C},\\
\label{S_EOM_3}
    \frac{\dd\sg_3^z}{\dd t} & = p\left( \langle \sg_3^z\rangle_{\rm eq} - \sg_3^z\right) - 2g\hat{C},
\end{align}
whose steady-state solution is
\begin{align}
\label{C_soln_3}
    \langle \hat{C}\rangle & = \frac{4gp}{p^2+ 8g^2}\langle  \sg^z_3\rangle_{\rm eq}, \\
    \label{S_soln_3}
    \langle  \sg^z_3\rangle & = \frac{p^2}{p^2+ 8g^2}\langle \sg^z_3\rangle_{\rm eq},
\end{align}
from which Eq.~\eqref{power_3QE} follows. 

The power fluctuations follow from $\dd \hat{W}^2/\dd t = gE_v\langle \hat{K}\rangle$, with $\hat{K}= \{\hat{W},\hat{C}\}$. Defining the Schr\"odinger-picture operator $\hat{\Omega} = \sg_3^z \hat{W}$, the relevant Heisenberg equations read as
\begin{align}
\label{K_EOM_3}
    \frac{\dd \hat{K}}{\dd t} & = 2g\left(2\hat{\Omega}-E_v\right) - \frac{p}{2}\hat{K},\\
\label{Omega_EOM_3}
     \frac{\dd \hat{\Omega}}{\dd t} & = p\left( \langle \sg_3^z\rangle_{\rm eq}\hat{W} - \hat{\Omega}\right) - g\hat{K}.
\end{align}
Eliminating $\hat{\Omega}$ and taking the long-time limit, we obtain
\begin{equation}
    \label{K_EOM_closed}
    \left[\frac{\dd^2}{\dd t^2}  + \frac{3p}{2}\frac{\dd}{\dd t}  + \frac{p^2}{2} + 4g^2 \right]\langle \hat{K}\rangle = 2g p E_v \left( 2\Gamma_3 t\langle \hat{Z}\rangle_{\rm eq}^2 + 1 \right).
\end{equation}
Seeking an asymptotic solution of the form $\langle \hat{K}\rangle = a + bt$, we deduce Eq.~\eqref{fluct_3QE}.

Finally, the energy currents can be computed using the master equation  describing the three qubits together with the load, as given in Appendix~\ref{app:3QE_ME}. Considering the equation of motion for $\langle \sg_j^z\rangle$ separately for $j=1,2,3$, and setting all derivatives to zero, one can show that a relation of the same form as Eq.~\eqref{J_solution} holds, but with $\hat{C}$ given in Eq.~\eqref{current_3QE}. Combining this with the solution~\eqref{C_soln_3}, Eq.~\eqref{ent_3QE} follows.

\bibliographystyle{apsrev4-1}
\bibliography{biblio}

\end{document}